\documentclass{iopart}
\usepackage{iopams}
\usepackage{amssymb,amstext,amsfonts,amsopn}
\usepackage{subeqn}
\usepackage{graphicx}

\def\indent#1{\begin{indented}\item[]{#1}\end{indented}}

\usepackage{color}

\begin{document}

\title%[]
{Static Structure of Active Brownian Hard Disks}
\author{N. de Macedo Biniossek$^1$, H. L\"owen$^1$,
  Th.~Voigtmann$^{1,2}$ and F. Smallenburg$^{1,3}$}
\address{%
  $^1$Institut f\"ur Theoretische Physik II: Weiche Materie,
  Heinrich Heine-Universit\"at D\"usseldorf,
  Universit\"atsstra\ss{}e 1, 40225 D\"usseldorf, Germany\\
  $^2$Institut f\"ur Materialphysik im Weltraum, Deutsches Zentrum
  f\"ur Luft- und Raumfahrt (DLR), 51170 K\"oln, Germany\\
  $^3$Laboratoire de Physique des Solides,  CNRS, Univ. Paris-Sud, Univ. Paris-Saclay, 91405 Orsay, France}
\eads{\mailto{thomas.voigtmann@dlr.de}}
\date\today

\def\eqref#1{(\ref{#1})}

\def\Fex{\ensuremath{F_\text{ex}}}
\def\vFex{\ensuremath{\vec F_\text{ex}}}
\def\Fexc{\ensuremath{F_\text{ex}^c}}
\def\kT{\ensuremath{k_\text{B}T}}

\begin{abstract}
We explore the changes in static structure of a two-dimensional system
of active Brownian particles (ABP) with hard-disk interactions,
using event-driven Brownian dynamics simulations. In particular, the effect of the self-propulsion velocity
and the rotational diffusivity on the orientationally-averaged
fluid structure factor is discussed. Typically activity increases
structural ordering and generates a structure factor peak at 
zero wave vector which is a precursor of motility-induced phase separation. Our results provide reference data to test future
statistical theories for the fluid structure of active Brownian systems. \\

\medskip

\noindent This manuscript was submitted for the special issue of the Journal of Physics: Condensed Matter associated with the Liquid Matter Conference 2017.
\end{abstract}

%\pacs{??}

\section{Introduction}

Active Brownian particles (ABP) are a widely used model system to study
the statistical physics of swimming micro-organisms \cite{Romanchuk_Schimansky-Geier_review_EPJST,Elgeti_Gompper_Winkler_review,ourRMP}.
In the ABP model, particles undergo Brownian translational and rotational
motion, and in addition possess an ``active'' mechanism of self-propulsion
along a fixed body axis. 
The simplest such model is that of active hard spheres (AHS), where particle
interactions are spherically symmetric and only enforce no-overlap conditions.
In particular, no direct interactions between the swimming directions of
AHS exist. Despite this simplification, AHS still show interesting
non-equilibrium phase behavior that allows to investigate many principles
of actively driven systems. They can also be realized to good approximation
in experiments on colloidal ``Janus'' particles \cite{buttinoni2012active,Howse2007,Palacci2014}.

For equilibrium ("passive") fluids, the basic quantity revealing the structural correlations 
in the disordered fluid state
is the structure factor $S(q)$, or alternatively the radial pair-distribution
function $g(r)$
\cite{HansenMcDonaldbook}. Since the early days of liquid state physics,
the structure factor 
has been measured by scattering experiments and computed by simulations and 
integral equation theory. In this respect the hard-sphere system 
(including its two-dimensional version of hard disks)
 has played a pivotal role to understand fluid structure and to test approximative theories.
The hard-sphere potential does not have an energy scale and therefore,
temperature is irrelevant in determining the phase behavior.
This allows to examine the role of structural correlations in fluids most
clearly.
%It is therefore a bit surprising that the
The
structural properties of active fluids have been much less
studied (but see \cite{Haertel} for a recent exception). % As in passive systems, 
This is a significant gap, because
the homogeneous active fluid can provide a clear-cut
testing ground to extend the well-established concepts of classical
statistical physics to regions far from thermal equilibrium.

In this contribution, we provide reference simulation data for the
static structure factor $S(q)$ and the pair distribution function $g(r)$
of active Brownian hard disks in two spatial dimensions. As stated above,
hard-sphere like interactions are the least arbitrary starting
point to describe sterical repulsion between particles, and in equilibrium
fluids, the approach to start from hard spheres and extend to other types
of interactions has been hugely successful. In active fluids, there might
be a subtle interplay between the softness of repulsion and active driving
\cite{Haertel}, so that it is important to establish the hard-sphere
reference case.
We restrict ourselves to two-dimensional systems for simplicity. To date,
most experiments on Janus colloids are done in quasi-2D settings, and
many simulation studies of the phase behavior of ABP have also been performed
in 2D. Brownian dynamics simulations of strict hard disks are not straightforward
 as already known from passive systems
(see e.g. \cite{Hinsen,LowenPRE1994,DeMichele}) so special care is needed for the algorithm used. Here 
we choose an event-driven scheme \cite{DeMichele} which is particularly designed 
to be efficient for Brownian hard disks. 

Sufficiently strong self propulsion induces clustering of ABP, so that for a
large range of densities, ABP systems evolve into inhomogeneous states
of very dense clusters separated by very dilute regions \cite{Marchetti,TailleurCatesReview2015}.
This phenomenon is called motility-induced phase separation (MIPS) as it
shares a number of qualitative features with the liquid--gas phase separation
known from equilibrium fluids.
MIPS has been studied in great detail for various
spherical ABP models with different interactions between the particles \cite{Bialke_review,Zoettl,Siebert},
since recently also including the hard-sphere case (using the same simulation
algorithm as ours) \cite{Levis}.
Here we deliberately restrict our attention to the homogeneous fluid state
outside the spinodal of MIPS.

\section{Methods and Techniques}

The active Brownian hard-disk system obeys the following equations of motion
for the positions $\vec r_j$ and the orientation angles $\theta_j$ of the
particles (relative to a fixed laboratory coordinate frame):
\begin{subequations}
\begin{eqnarray}
  d\vec r_j&=&\sqrt{2D_t}\,d\vec W_j+v_0\vec e(\theta_j)\,dt\,,
  \qquad|\vec r_j-\vec r_k|\ge\sigma\;\forall j,k\,,
  \\
  d\theta_j&=&\sqrt{2D_r}\,dW^\theta_j\,.
\end{eqnarray}
\end{subequations}
Here, $j=1,\ldots N$ labels the particles. Brownian translational and
rotational diffusion is described by uncorrelated Wiener processes
$d\vec W_j$ and $dW^\theta_j$; their amplitude is given by the
translational diffusion coefficient $D_t$ and the rotational diffusion
coefficient $D_r$. Self propulsion is modeled by a fixed swimming speed
$v_0$ along the particles orientation, $\vec e(\theta)=(\cos\theta,
\sin\theta)^T$. The hard-sphere interactions translate into no-overlap
conditions $|\vec r_j-\vec r_k|\ge\sigma$ for all particle pairs.

The hard-core diameter $\sigma$ and the translational diffusion coefficient $D_t$
are used to set the units of length and time in the following.
There remain three parameters to specify the state of the system:
the density $n=N/V$, expressed as a dimensionless packing fraction
$\eta=(\pi/4)n\sigma^2$, the self-propulsion velocity $v_0$ (in units
of $D_t/\sigma$), and the rotational diffusion coefficient $D_r$
(in units of $D_t/\sigma^2$).
For three-dimensional passive hard spheres, the Stokes-Einstein relation
fixes $D_r=3D_t/\sigma^2$, assuming Stokes flow in the solvent and stick
boundary conditions on the particle surface. However, for active systems,
the effective rate of change of particle orientations may be significantly
different from this passive value, depending on the swimming mechanism.
We will therefore fix $D_r=1$ as a reference case for most simulations,
and also explore the effect of changes in $D_r$, i.e., changes in the
persistence of active motion.

Given the particle positions, one obtains the static structure factor,
\begin{equation}\label{eq:sq}
  S(q)=\frac1N\left\langle\sum_{j,k=1}^N e^{-i\vec q\cdot(\vec r_j-\vec r_k)}
  \right\rangle\,,
\end{equation}
where the angular brackets denote an average over the non-equilibrium
stationary state, and $\vec q$ is the wave vector of the density fluctuations
that are probed by $S(q)$. Note that in the homogeneous, isotropic fluid,
$S(q)$ depends on $\vec q$ only through its magnitude $q=|\vec q|$.
The static structure factor is intimately related to the radial distribution
function $g(r)$ defined by
\begin{equation}\label{eq:gr}
  g(r)=\frac1{nN}\left\langle\sum_{j\neq k}\delta(\vec r-(\vec r_j-\vec r_k))
  \right\rangle\,,
\end{equation}
which quantifies the probability density for finding a particle at distance
$r$ from a given particle, irrespective of their orientations.
Again, $g(r)$ quantifies structural properties of the isotropic homogeneous
fluid, although for ABP systems, the angle-resolved pair distribution
function provides further information \cite{Haertel}.

Simulations were carried out using an event-driven Brownian dynamics (ED-BD)
algorithm \cite{DeMichele}. In the ED-BD simulation, a fixed ``Brownian''
time step $\Delta t$ is introduced, and at each time step, Gaussian trial
displacements $\Delta\vec r_i$ and angle increments $\Delta\theta_i$
are drawn for all the particles. The self-propulsion term is included
by drawing the $\Delta\vec r_i$ from appropriately shifted Gaussians.
To propagate the system to the next time steps, the trial displacements
have to be modified to avoid particle overlaps. This is done by assigning
to the particles pseudo-velocities $\vec v_i=\Delta\vec r_i/\Delta t$,
and by performing event-driven molecular dynamics using these pseudo-velocities
in the time interval $[t,t+\Delta t]$.
This way, the ED-BD algorithm guarantees no-overlap conditions
at any time, and thus incorporates hard-sphere interactions exactly.

In the passive case, the ED-BD algorithm has been shown to accurately
describe the Brownian motion of hard spheres, if the time step is reasonably
small, $\Delta t<0.1\,\sigma^2/D_t$, say \cite{DeMichele}. Note however
that for strong self-propulsion and/or fast reorientational diffusion,
significantly smaller time steps may be required.
The extension
to active particles has been used to study glassy dynamics of dense
AHS systems \cite{Ni2013}, and more recently also MIPS \cite{Levis}.

We performed runs with $N=1000$ and $N=5000$ particles to estimate
finite-size effects. These are found to be small for the structural
quantities we study for state points sufficiently far from the spinodal of MIPS.
Results are shown for the $N=5000$ system. The Brownian time step 
was chosen to be $\Delta t = 0.001\,\sigma^2/D_t$, and some runs with
$\Delta t=0.01\,\sigma^2/D_t$ were performed to confirm that no significant
finite-time step effects remain in $S(q)$.
From individual runs with a specific choice of parameters, after an initial
transient time of $t_i=333\,\sigma^2/D_t$ to reach a stationary state,
133 configurations in $[t_i,t_f]$ with $t_f=1000\,\sigma^2/D_t$ were
stored and analyzed to obtain $S(q)$ and $g(r)$. To obtain $S(q)$,
a set of 5000 $\vec q$-vectors compatible with periodic boundary conditions
were used to evaluate Eq.~\eqref{eq:sq}, which were afterwards binned
according to $|\vec q|$. For $g(r)$, bins of width $0.01\,\sigma$ were used
in evaluating Eq.~\eqref{eq:gr}.

\section{Results}

\subsection{Passive Hard Disks}

\begin{figure}
\indent{\includegraphics[width=\linewidth]{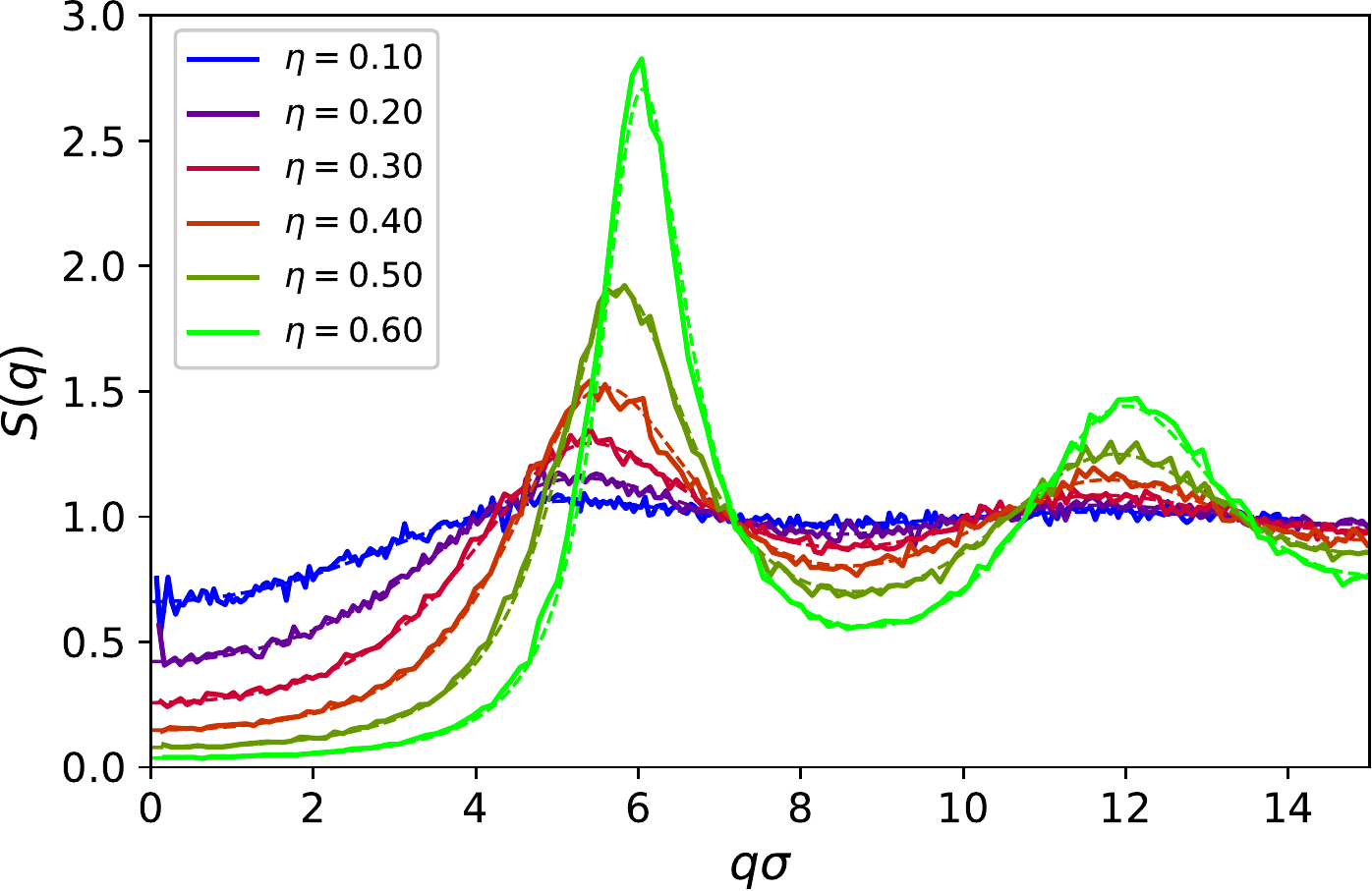}}
\caption{\label{fig:sq_passive}
  Static structure factor $S(q)$ of passive hard-disk systems with
  packing fractions $\eta=0.1$, $0.2$, $0.3$, $0.4$, $0.5$, and $0.6$
  (top to bottom around $q\sigma=2$).
  Solid lines: results from ED-BD simulations.
  Dashed lines represent the Baus-Colot expression for $S(q)$.
}
\end{figure}

We begin by recalling the features of the passive reference system.
The only relevant parameter in the passive hard-disk system is the packing
fraction $\eta$.
Hard disks are known to order at high densities, transforming first
from the fluid to a hexatic phase, and later to a crystalline phase.
The nature and phase-transition boundaries of the fluid--hexatic and
the hexatic--solid transitions have only recently become clear in large-scale
simulations \cite{Bernard2011} and experiments \cite{Thorneywork2017}:
in the regime $0.700\le\eta\le0.716$, coexistence between a fluid and
the hexatic phase was found. There is a further continuous transition
to a solid at $\eta\simeq0.720$.
In the following, we will restrict the discussion to packing fractions
$\eta\le0.7$.

The static structure factor of the fluid displays the standard features
known from simple fluids (Fig.~\ref{fig:sq_passive}): with increasing
packing fraction, intermediate-range order in the fluid becomes more
pronounced, and this gives rise to damped oscillations in $S(q)$ that
become increasingly pronounced. The ordering is reflected in a pronounced
first peak of $S(q)$ at $q_*\approx6/\sigma$ (for $\eta=0.6$).
The position of this peak
(or rather, the period of the oscillations) reflects a typical interparticle
distance. The sharpness of the peak is an indicator for how pronounced
ordering is.

For 3D hard spheres, several well known approximation schemes exist
for $S(q)$.
For example, a widely used analytical, albeit approximate, expression is the
Percus-Yevick (PY) structure factor \cite{HansenMcDonaldbook}.
In 2D, no closed analytical form of the PY approximation for $S(q)$
is known.
An empirical expression has been proposed by Baus and Colot \cite{Baus}.
The Baus-Colot expression provides an excellent description of the
data, as shown by the dashed lines in Fig.~\ref{fig:sq_passive}.

\subsection{Active Hard Disks}

\begin{figure}
\indent{\includegraphics[width=\linewidth]{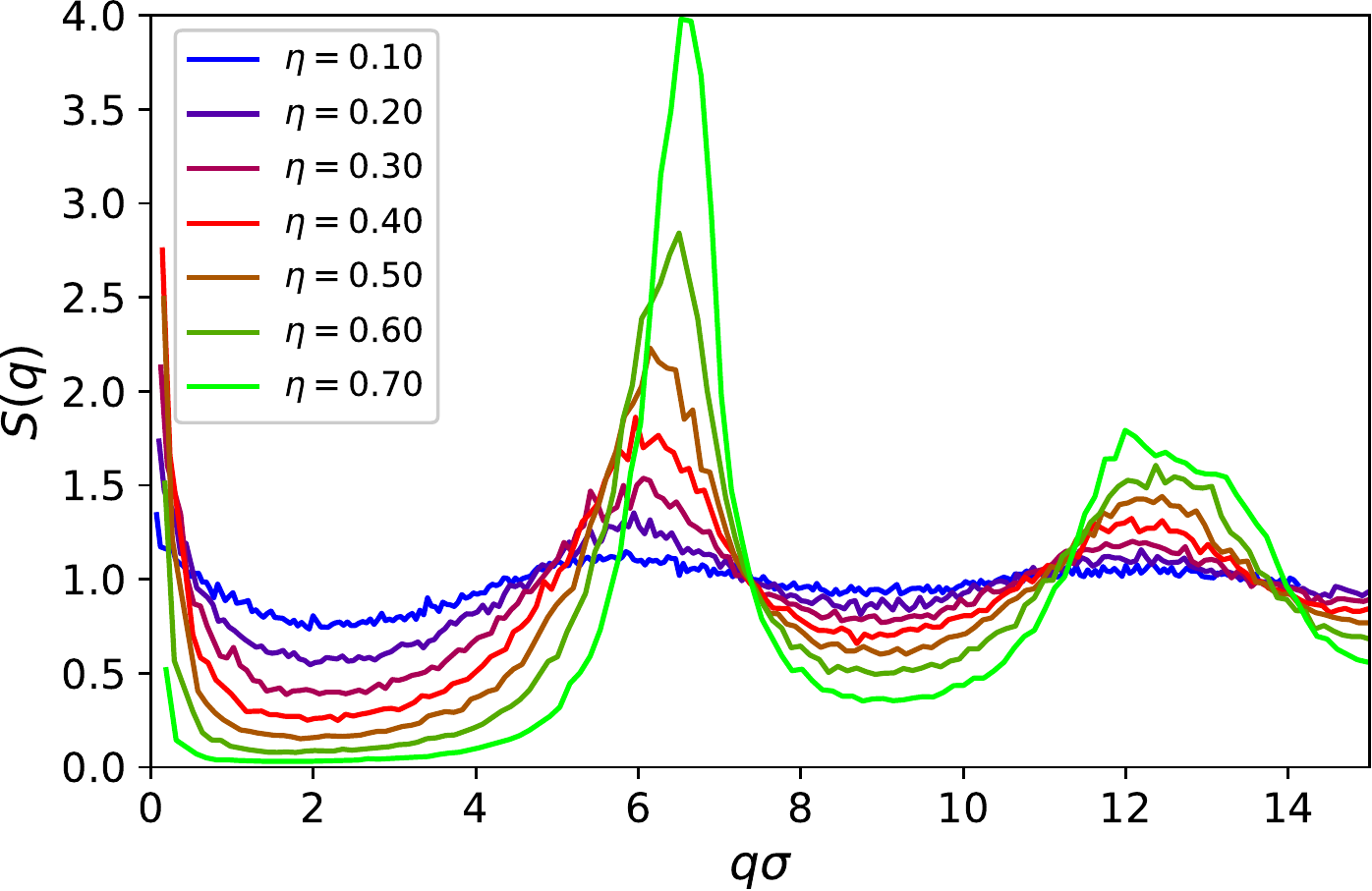}}
\caption{\label{fig:sq_lowv0}
  Static structure factor $S(q)$ of active Brownian hard disks with
  self-propulsion velocity $v_0=10\,D_t/\sigma$ and rotational diffusion
  coefficient $D_r=1\,D_t/\sigma^2$, for packing fractions
  $\eta=0.1$, $0.2$, $0.3$, $0.4$, $0.5$, $0.6$, and $0.7$.
}
\end{figure}

We now turn to the discussion of the active hard-disk system with rotational
diffusivity $D_r=1$.
For self-propulsion velocities $v_0$ that are below the onset of
motility-induced phase separation (estimated to be around $v_0=12$ in
the present system \cite{Levis}), a similar evolution of $S(q)$
with increasing packing fraction is seen as in the passive system.
This case is shown in Fig.~\ref{fig:sq_lowv0} for $v_0=10$.

In particular the evolution of the first and second peaks in $S(q)$ does
not differ qualitatively from the one in the passive system. For
$\eta=0.7$, the second peak around $q\sigma=12$ exhibits an asymmetric
shape, which we interpret as a precursor of incipient ordering.
It is known that two-dimensional ABP systems crystallize at high densities
(seen, for example, in active hard-core Yukawa systems \cite{Bialke2012}).
For the hard-disk case discussed here, the passive system displays
much stronger signatures of ordering in $S(q)$ at $\eta=0.7$ than the
active system does. This is consistent with the expectation that -- at
least for self-propulsion velocities small enough to prevent MIPS -- the
ordering transition sets in at higher densities in the presence of
active motion.

The main difference of the active $S(q)$ to the passive one
is in the low-$q$ behavior.
At densities comparable to the critical
density of MIPS, a strong increase of $S(q\to0)$ is seen in the active system.
This is
the signature of impending phase separation that is expected from the analogy
with equilibrium systems.
Precursors of this low-$q$ increase are seen at all densities shown
in Fig.~\ref{fig:sq_lowv0}.

The structure factors shown in Fig.~\ref{fig:sq_lowv0} are exemplary for
the active hard-disk fluid. At higher $v_0$, only the low-density regime
remains, because MIPS sets in; our simulations confirm that phase separation
prevails for large $v_0$ up to very high densities. It is therefore,
at least for $D_r\approx1$,
not possible to prepare a homogeneous monodisperse hard-disk fluid for
large $v_0$ and large $\eta$.

\begin{figure}
\indent{\begin{tabular}{ll}
a) & b) \\
 \includegraphics[height=0.42\linewidth]{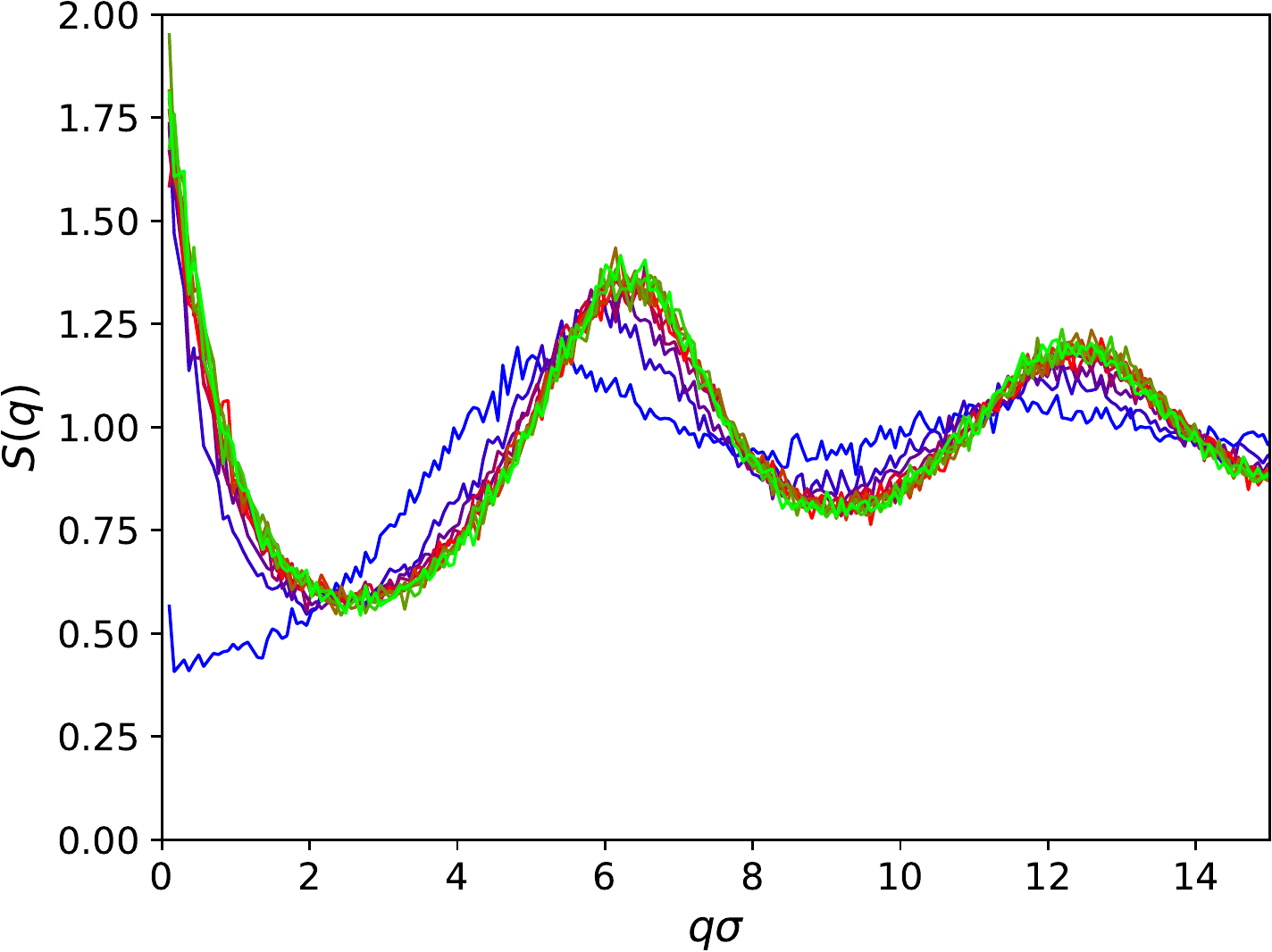} &
\includegraphics[height=0.42\linewidth]{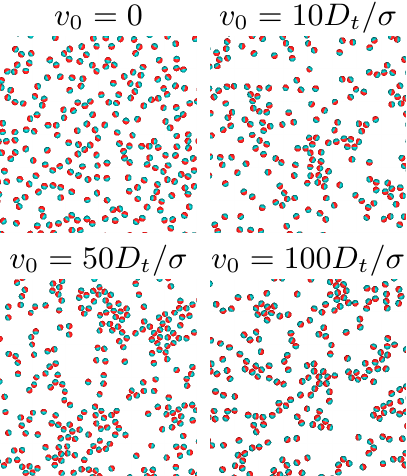} 
\end{tabular}}
\caption{\label{fig:sq_lowdens}
  a) Low density fluid regime: static structure factor $S(q)$ for active hard disks with $D_r=1\,D_t/\sigma^2$
  and at packing fraction $\eta=0.2$. Different curves corresond to
  different self-propulsion velocity: $v_0=0\,D_t/\sigma$ to $v_0=100\,D_t/\sigma$ increasing in
  steps of $10$ (blue to green). b) Simulation snapshots at the indicated self-propulsion velocities. The self-propulsion
  of each particle is directed towards its red side. Note that the total simulation box is significantly larger than the
  region depicted here.
}
\end{figure}

To understand the effect of activity on the static structure of the fluid,
it is instructive to discuss cuts in parameter space where the packing
fraction is fixed. Increasing the self-propulsion velocity $v_0$
for a low-density system ($\eta=0.2$ shown in Fig.~\ref{fig:sq_lowdens}a)
causes two prominent changes: first, the main peak of $S(q)$ shifts
to larger $q$ and increases in amplitude (and the subsequent peaks
undergo a similar change).
The increase of the peaks in $S(q)$ signals that structural order in the
more active fluid is more pronounced, while the shift to larger $q$
indicates that the average particle distance is reduced. This is consistent with 
a visual inspection of the simulation snapshots illustrated in Fig.~\ref{fig:sq_lowdens}b, which
shows the formation of local areas with both higher ordering and density.

Second, the low-$q$ region of $S(q)$
increases with increasing self-propulsion speed.
This latter effect is attributed to the emergence of
a phase-separated region, as discussed above.
If one defines the isothermal compressibility $\kappa_T$
of the active fluid system in terms of the particle-number fluctuations
by extending the well-known equilibrium relation,
$\kappa_T=S(q\to0)/n k_BT$ (where $k_BT$ is the thermal energy needed to
define an energy scale for the compressibility), to the non-equilibrium
stationary state, the active fluid is much more compressible than the
passive one.

Interestingly, the $S(q)$ appear to approach a limiting curve for large $v_0$:
in Fig.~\ref{fig:sq_lowdens}a, the strongest change is seen upon increasing
$v_0$ from zero to about 10, while a further increase by a factor of $10$ (up
to $v_0=100$) only causes small further changes in the average fluid structure.
Also the MIPS spinodal is nearly vertical in the $v_0$-vs-$\eta$ plane
\cite{Levis},
indicating that there is a regime where self-propulsion effects saturate.
The shift in the peak positions of $S(q)$ indicate that the average
particle distance decreases from about $2\pi/5\,\sigma\approx1.26\,\sigma$ to
about $2\pi/6\,\sigma\approx1.05\sigma$. Hence, the saturation may stem from the
fact that for true hard disks, a further increase in activity cannot
cause particles to stay closer than $\sigma$ on average.
The saturation effect may thus be masked for soft spheres.

\begin{figure}
\indent{\includegraphics[width=\linewidth]{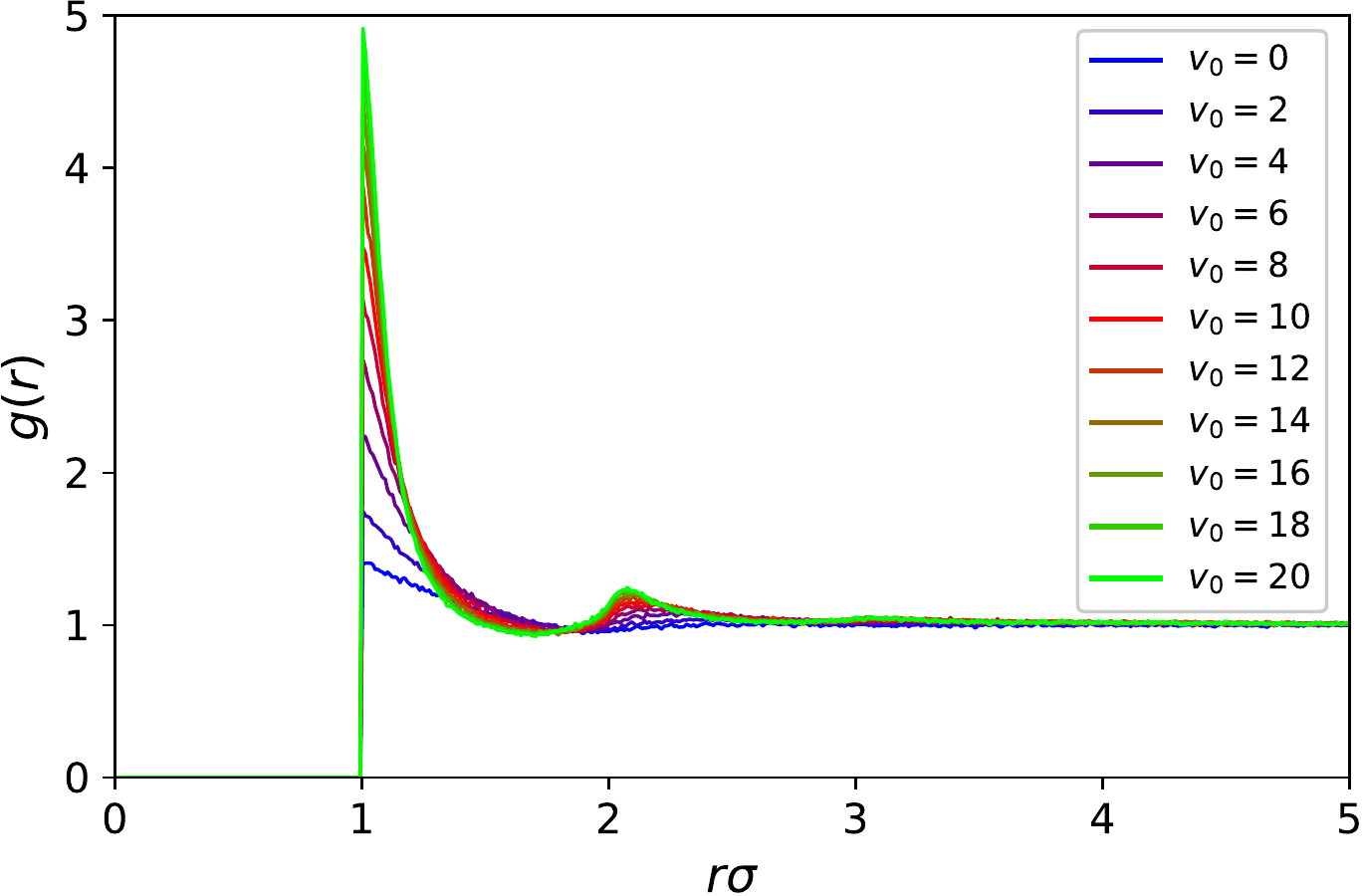}}
\caption{\label{fig:gr_lowdens}
  Radial distribution function $g(r)$ of an active Brownian hard-disk system
  at packing fraction $\eta=0.2$, for self-propulsion velocities $v_0
  \in[0,20]$ (in steps of two, from bottom to top at $r=1^+$).
}
\end{figure}

The radial distribution function $g(r)$ provides more intuitive
information on the average structure. At the low density discussed here,
$\eta=0.2$, the equilibrium $g(r)$ is relatively featureless: it shows
a weak enhancement over the ideal-gas value at particle contact, but quickly
decays to unity for larger $r$.
Increasing $v_0$ for the low-density
system leads to a strong enhancement of $g(r)$ near contact, see
Fig.~\ref{fig:gr_lowdens}. This is in line with the interpretation that
the average particle distance decreases with increasing self-propulsion
velocity.

At sufficiently large $v_0$ there emerges a second peak around $r=2\sigma$.
Hence, active particles form small transient clusters. However, no
pronounced peaks are seen at $r\approx n\sigma$ with $n>2$. Thus,
the system does not (yet) form large clusters with a statistically
significant probability, and it also does not form pronounced intermediate-range
order.
%The resulting $g(r)$ resembles the one of a
%mixture of monodisperse hard disks with a small fraction of dimers.

The changes in $g(r)$ that are visible in Fig.~\ref{fig:gr_lowdens} can
be contrasted to those found in ABP systems with soft
interactions \cite{Brader15}. There, a comparable change of $v_0$ only
led to a relatively mild change in $g(r)$, and the main effect was a shift
of the nearest-neighbour peak to smaller distances.

An appealing concept is to map activity onto an effective interaction between
the particles. Quite robustly for a number of different ABP models,
activity induces effective attractions that become increasingly strong when
$v_0$ is increased \cite{Brader15,Ginot,Mani,Wittmann}.
For soft-sphere ABP, the interaction range was found to be around $20\%$ of a
particle diameter at moderate densities \cite{Brader15}.

To describe the effects of attractions in passive colloidal systems,
the square-well system (SWS) is a canonical starting point \cite{Bolhuis1994}. In this model,
a hard-sphere repulsion is supplemented by an attraction of fixed strength
$\Gamma$ and range $\delta\sigma$. For the three-dimensional SWS, the static
structure factor $S(q)$ can be obtained analytically within the mean-spherical
approximation for $\delta$ not too large \cite{Dawson2001}. At low density,
the SWS-$S(q)$ displays a change upon increasing the attraction
strength $\Gamma$ that is similar to the one seen for the low-density
AHS in Fig.~\ref{fig:sq_lowdens}: increasing attraction causes increased
structural order, and a shift of the average particle separation to lower
distances. This confirms earlier findings that activity can be mapped
to an effective attraction in describing the fluid structure \cite{Ginot}. The mapping
$v_0\leftrightarrow\Gamma$
is however quite nonlinear, because the saturation effect we find for the AHS
upon increasing $v_0$ is not found in the SWS upon increasing $\Gamma$.

\begin{figure}
\indent{\parbox[t]{\linewidth}{
  \parbox[t]{.495\linewidth}{{\includegraphics[width=\linewidth]{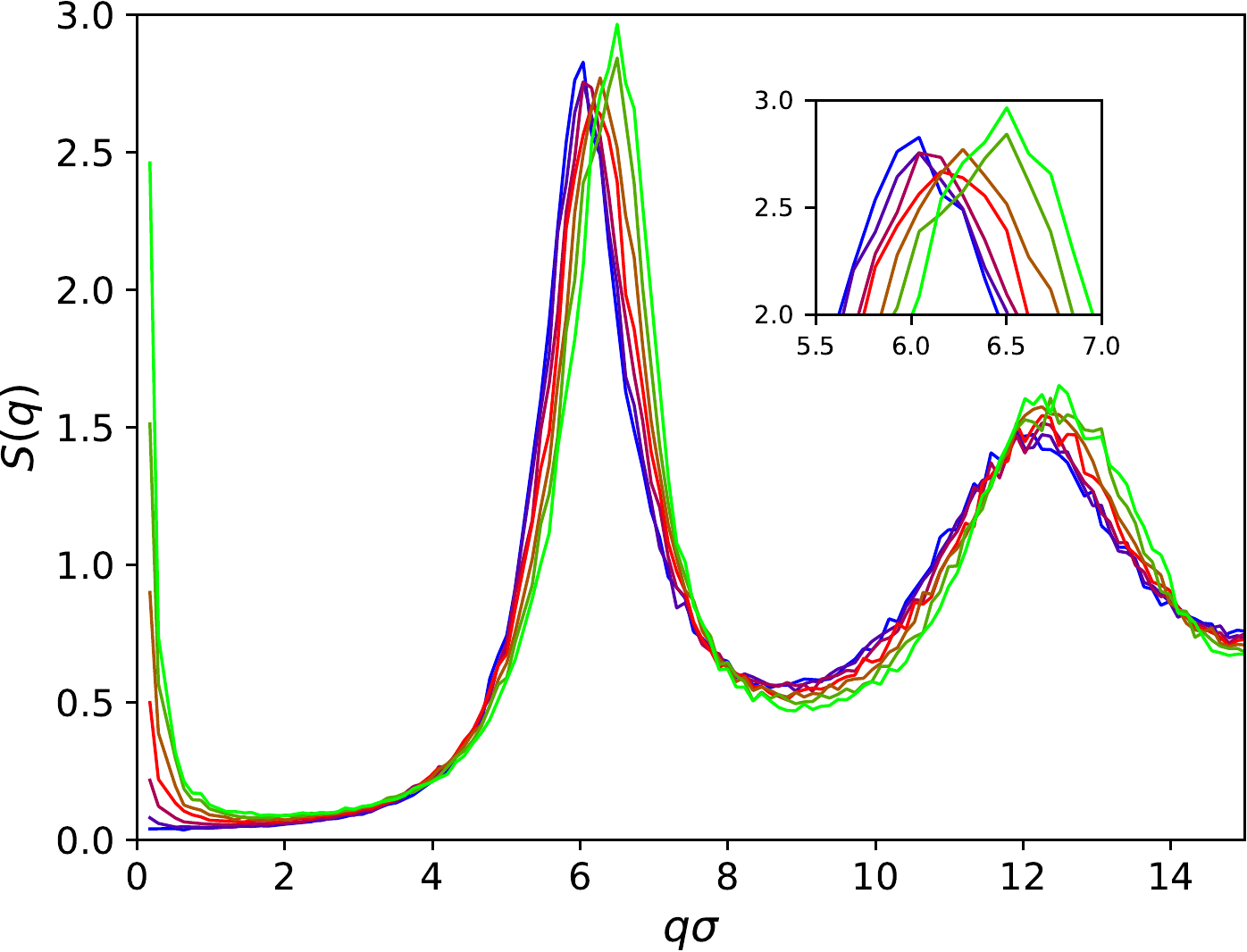}}}\hfill%
  \parbox[t]{.495\linewidth}{{\includegraphics[width=\linewidth]{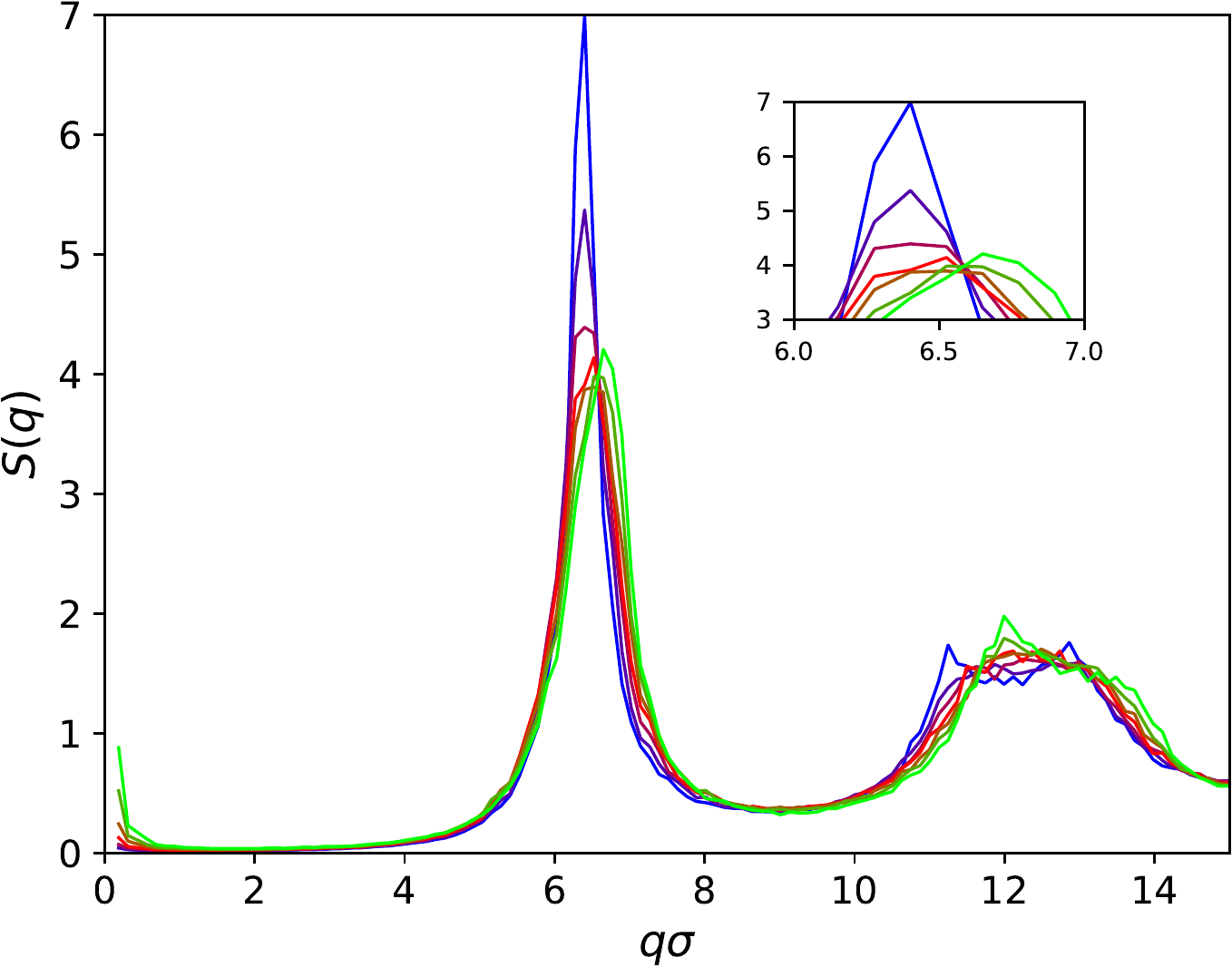}}}\\
  \parbox[t]{.495\linewidth}{{\includegraphics[width=\linewidth]{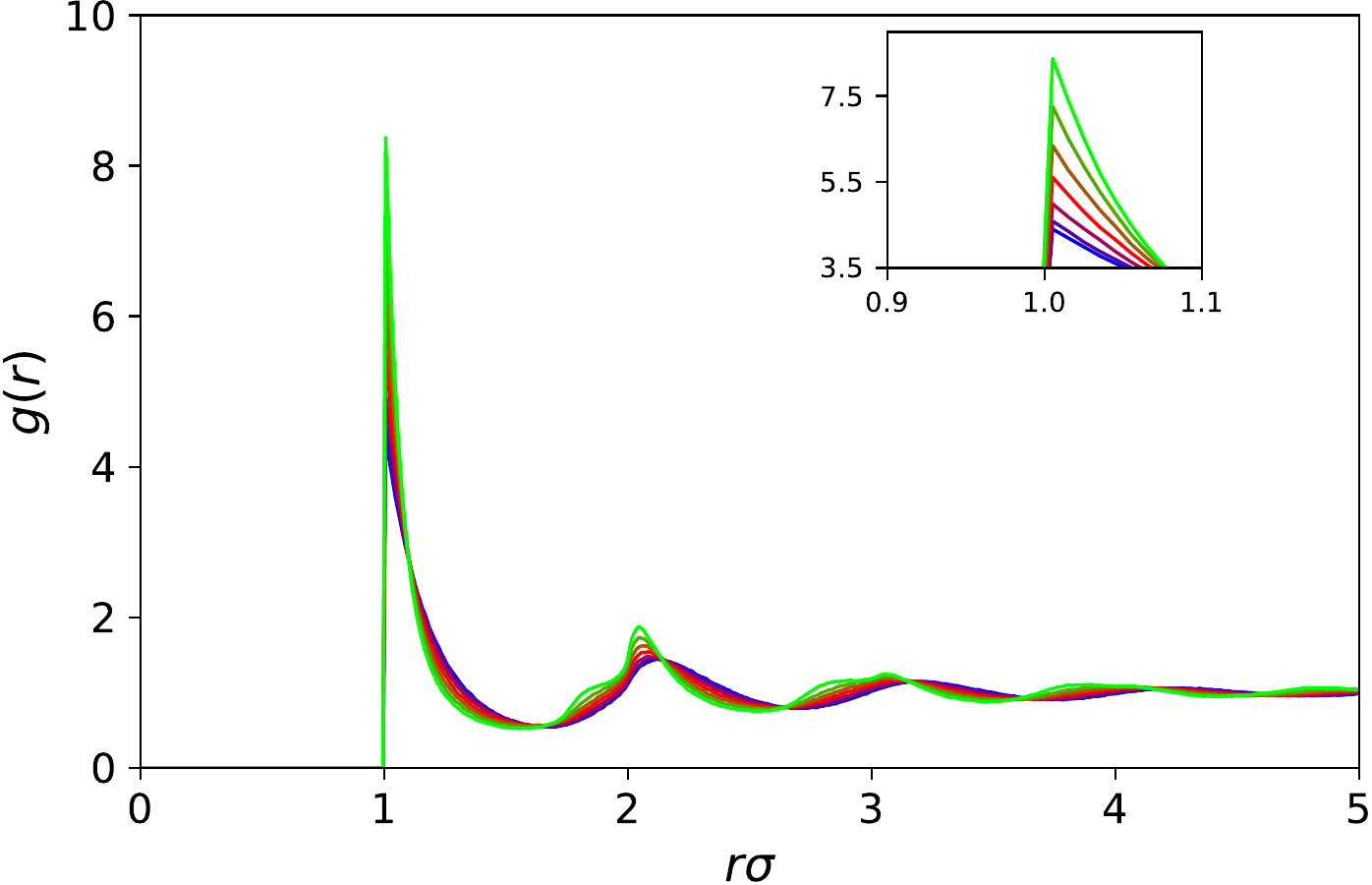}}}\hfill%
  \parbox[t]{.495\linewidth}{{\includegraphics[width=\linewidth]{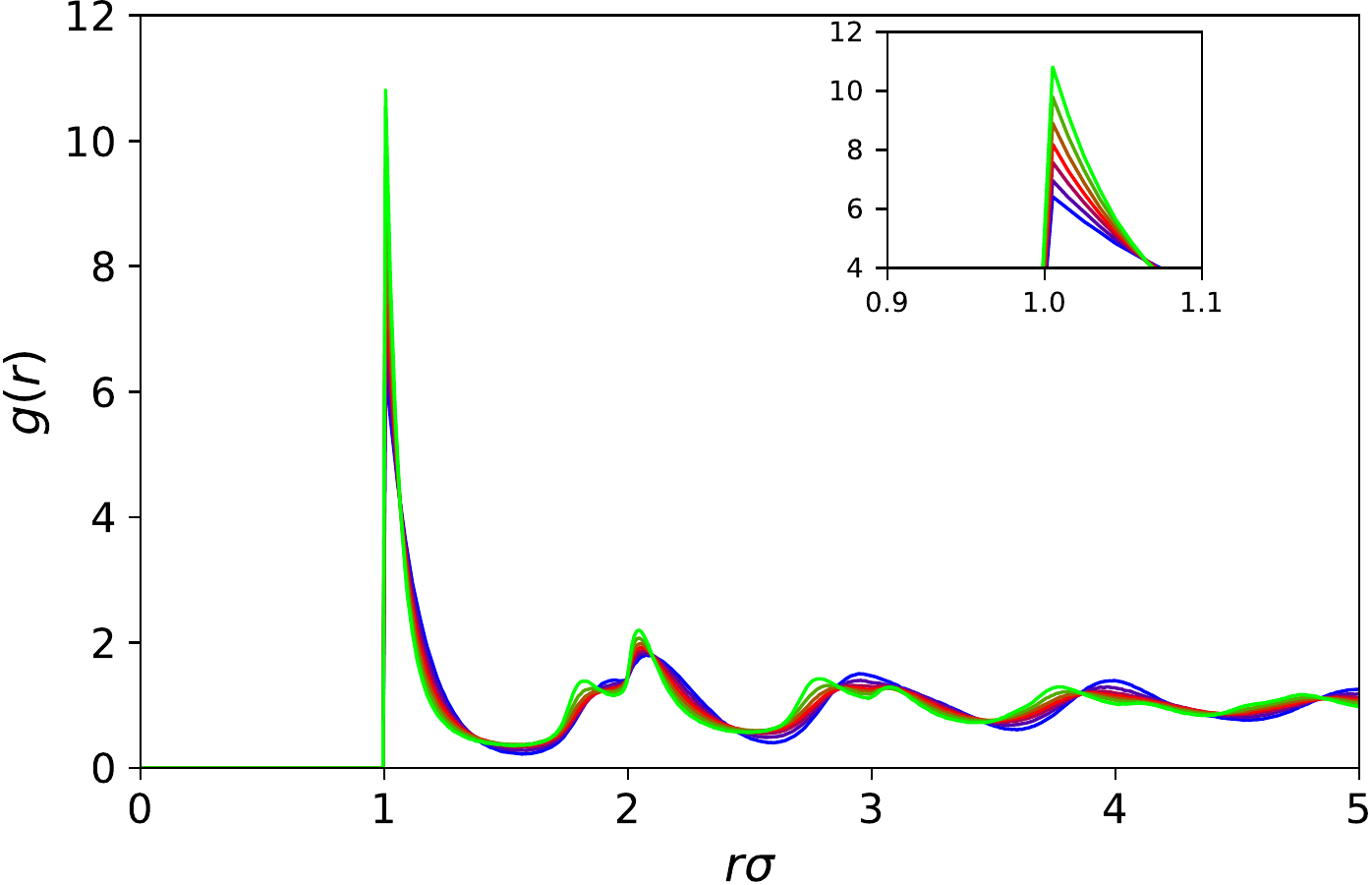}}}
  \parbox[t]{.495\linewidth}{{\includegraphics[width=0.8\linewidth]{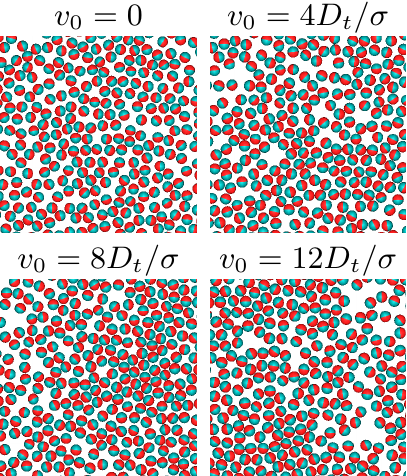}}}\hfill%
  \parbox[t]{.495\linewidth}{{\includegraphics[width=0.8\linewidth]{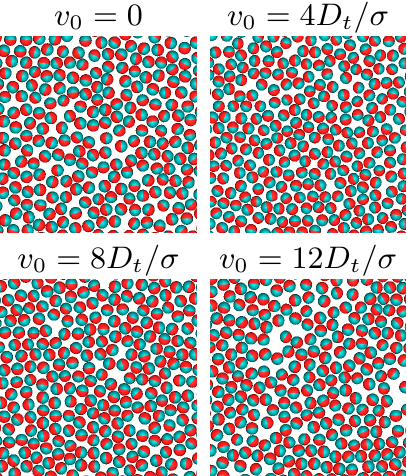}}}
}}
\caption{\label{fig:sq_highdens}
  High-density fluid regime: static structure factor $S(q)$ (top row), 
  radial distribution function $g(r)$ (middle row), and snapshots (bottom row) of active hard disks
  with packing fraction $\eta=0.6$ (left column) respectively
  $\eta=0.7$ (right column). Different curves correspond to different
  self-propulsion speeds $v_0=0$, $2$, $4$, $6$, $8$, $10$, $12$
  (from blue to green; curves ordered by increasing
  first-peak position in $S(q)$ from left to right).
}
\end{figure}

We now turn to the high-density fluid, limited to small enough $v_0$
so that the system remains homogeneous. As the phase diagram confirms
\cite{Levis},
there opens a small window where the system is not yet crystallized
and not yet phase-separated. We discuss two packing fractions for this
case, $\eta=0.6$ and $\eta=0.7$. The latter case represents the upper
end of the equilibrium fluid regime in passive hard disks, and the
structure functions for this system already show precursors of a
phase transition to the hexatic phase.

The $S(q)$ at high densities, Fig.~\ref{fig:sq_highdens},
demonstrate an interesting trend upon increasing $v_0$ that is
absent at low densities: while the main peak of $S(q)$ monotonically
shifts to the right with increasing $v_0$, it first weakens, and then
increases with $v_0$.
The corresponding effect in $g(r)$ is an interplay between a sharpening of
the contact-value peak and an increasing depletion of the region between
the first neighboring shells. However, this structural
change is too subtle to be clearly visible in the simulation snapshots.
The initial decrease in the amplitude of $S(q)$
is much more pronounced at $\eta=0.7$, but already noticable at $\eta=0.6$,
where precursors of hexatic ordering are not yet obvious.
We therefore attribute the non-monotonic change to a genuine change in
the way activity influences the fluid structure, and not to the vicinity of
a phase transition.

The non-monotonic trend revealed in Fig.~\ref{fig:sq_highdens} indicates
that the effect of activity in the high-density hard-disk system is
twofold: first, activity reduces ordering in the dense system.
This is also expected from
simulation studies of glassy dynamics, where a shift of the glass transition
to higher densities with increasing activity was seen \cite{Ni2013}. Similar
trends are confirmed for various other model systems \cite{Berthier2013,Bi2016}
and predicted by theory \cite{mct}. Such a ``fluidization'' of the system
is usually associated with a weakening of the peak amplitudes in $S(q)$.
It is reminiscent of the reentrant melting of glass-forming systems with
short-ranged attraction, where a similar decrease of peak height in $S(q)$
with increasing attraction strength describes the structural changes
of the system \cite{Dawson2001}.
Second, stronger activity in the high-density system restores the effects
that also prevail at lower densities. Here, activity favors structural
order. For the case we study here, the crossover between the two
effects occurs around $v_0=5\,D_t/\sigma$, slightly depending on the packing fraction.

\begin{figure}
\indent{\parbox[t]{\linewidth}{
  \parbox[t]{.495\linewidth}{{\includegraphics[width=\linewidth]{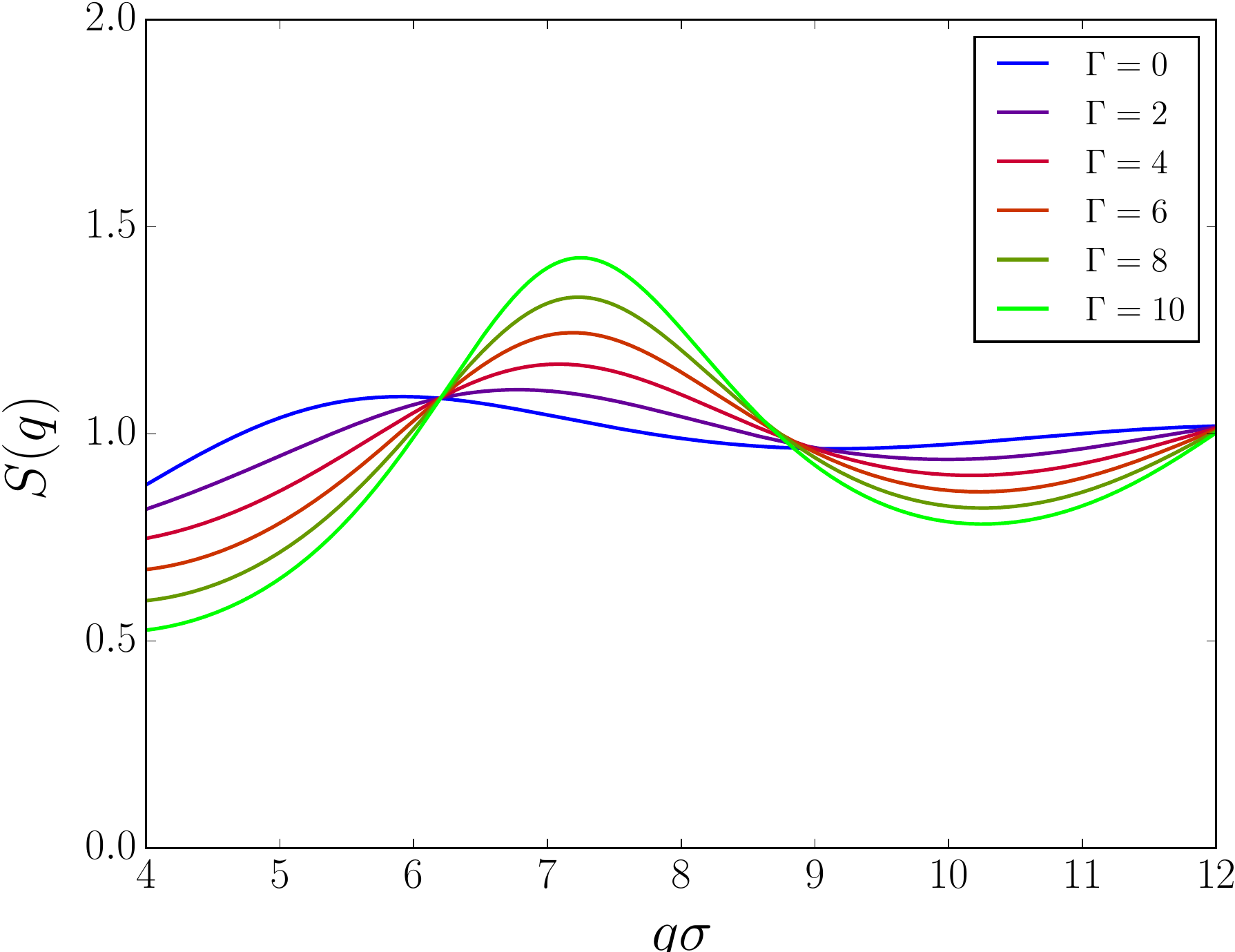}}}\hfill%
  \parbox[t]{.495\linewidth}{{\includegraphics[width=\linewidth]{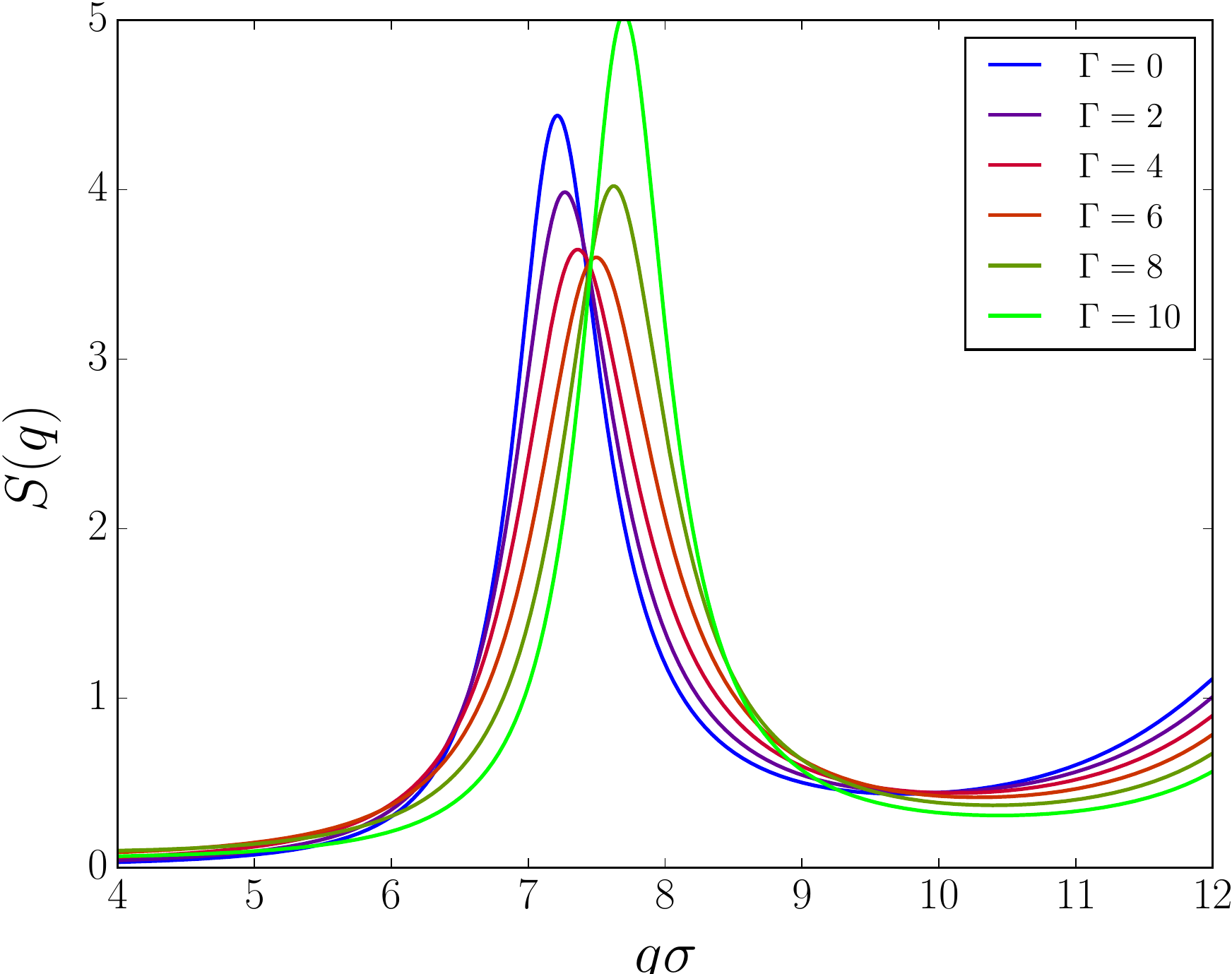}}}
}}
\caption{\label{fig:swssq}
  Static structure factor $S(q)$ of a three-dimensional square-well system
  with short-ranged attraction (range $0.1\sigma$), as a function of
  attraction strength ($\Gamma=0$, $2$, $4$, $6$, $8$, and $10$; blue to green),
  for moderate density (left panel; three-dimensional packing fraction
  $\varphi=0.1$) and for high density (right panel; $\varphi=0.55$).
}
\end{figure}

Again, the comparison with the passive square-well system allows to understand
the qualitative mapping of activity to effective interactions.
Specifically, the non-monotonic change in the first peak of $S(q)$ allows
to estimate the range of the effective attraction. For the SWS with an
attraction range of, say, $\delta=20\%$,
the same trend is found at large densities
as it was also found in the low-density state: increasing attraction strength
$\Gamma$ increases the peak height and shifts its position to larger $q$
(recall Fig.~\ref{fig:sq_lowdens}). The situation changes however, if one
considers short-ranged attractions. For, say, $\delta=10\%$, the dense SWS
reproduces a decrease in peak height with increasing $\Gamma$ up to
about $\Gamma=5\,k_BT$, followed by an increase upon further increasing
$\Gamma$. (To exemplify this, Fig.~\ref{fig:swssq} shows the corresponding
$S(q)$ for a three-dimensional $S(q)$, evaluated in the mean-spherical
approximation \cite{Dawson2001}.) 
From this similarity in trends one may conclude that indeed, activity in the AHS
can be mapped to an effective attraction, of a range around $10\%$.
The SWS also offers a physical explanation for the different behavior
of $S(q)$ at high densities as compared to low densities: at $\eta\approx0.6$,
the average particle separation is on the order of $10\%$ of a particle
diameter.
If the attraction range and the interparticle separation are comparable,
the effect of increasing attraction strength is to increase disorder,
because some particles will be bounded more strongly, while others can explore
a larger configuration-space volume. Only if the attraction is sufficiently
strong, the energetic increase in order offsets the entropic decrease.

\subsection{Influence of Rotational Persistence}

Activity induces a coupling between $D_r$ and the translational
evolution. Thus, $v_0$ and $D_r$ are both relevant parameters for the
active fluid.
One way to describe ABP that is successful in the low-density limit is
by assigning to the ABP system an effective temperature (or pressure)
\cite{Takatori2014},
based on the notion that a single ABP over long time and length scales
undergoes diffusion with a renormalized diffusivity that in two spatial
dimensions reads
$D_\text{eff}=D_t+(1/2)v_0^2/D_r$.
Here, only a specific combination of $v_0$ and $D_r$ enters.
There are indications that the high-density dynamics of ABP on the contrary
depends on
both these parameters separately \cite{mct,Ni2013,Bi2016},
because at high densities the average distance between ineracting particles
is easily shorter than the persistence length associated to the swimming,
$\ell_p=v_0/D_r$.
It is therefore interesting to see the effect that varying both $v_0$ and $D_r$
has on the stationary static structure.

The relevant dynamical rates of the AHS system thus are:
the rate of passive Brownian diffusion, $\tau_0^{-1}=D_t/\sigma^2$,
the rate of self-propelled motion, $\tau_v^{-1}=v_0/\sigma$,
and the rate of active effective diffusion, $\tau_a^{-1}=v_0\ell_p/\sigma^2$.
From these rates, three dimensionless parameters can be formed to quantify
the amount of ``activity'' in the system. (They are ratios of rates, and
thus called ``P\'eclet'' numbers in analogy to hydrodynamic theory.)
A natural choice from the equations of motion is the
translational P\'eclet number that quantifies the rate of self propulsion
in relation to passive diffusion,
${Pe}_t=\tau_v^{-1}/\tau_0^{-1}=v_0\sigma/D_t$.
The low-density theory suggests to use the rate of active over that of
passive diffusion,
${Pe}=\tau_a^{-1}/\tau_0^{-1}=v_0^2/D_rD_t$;
at high densities one may expect the persistence length to play a crucial
role, and thus the ratio of self-propelled motion relative to active
difusion,
${Pe}_r=\tau_v^{-1}/\tau_a^{-1}$. Note that
${Pe}_r^{-1}=\ell_p/\sigma$ is sometimes also referred to by the symbol
${Pe}$ in the context of MIPS; we stick to the notation introduced in
\cite{Takatori2014} to avoid confusion.

\begin{figure}
\indent{\parbox[t]{\linewidth}{
  \parbox[t]{.325\linewidth}{{\includegraphics[width=\linewidth]{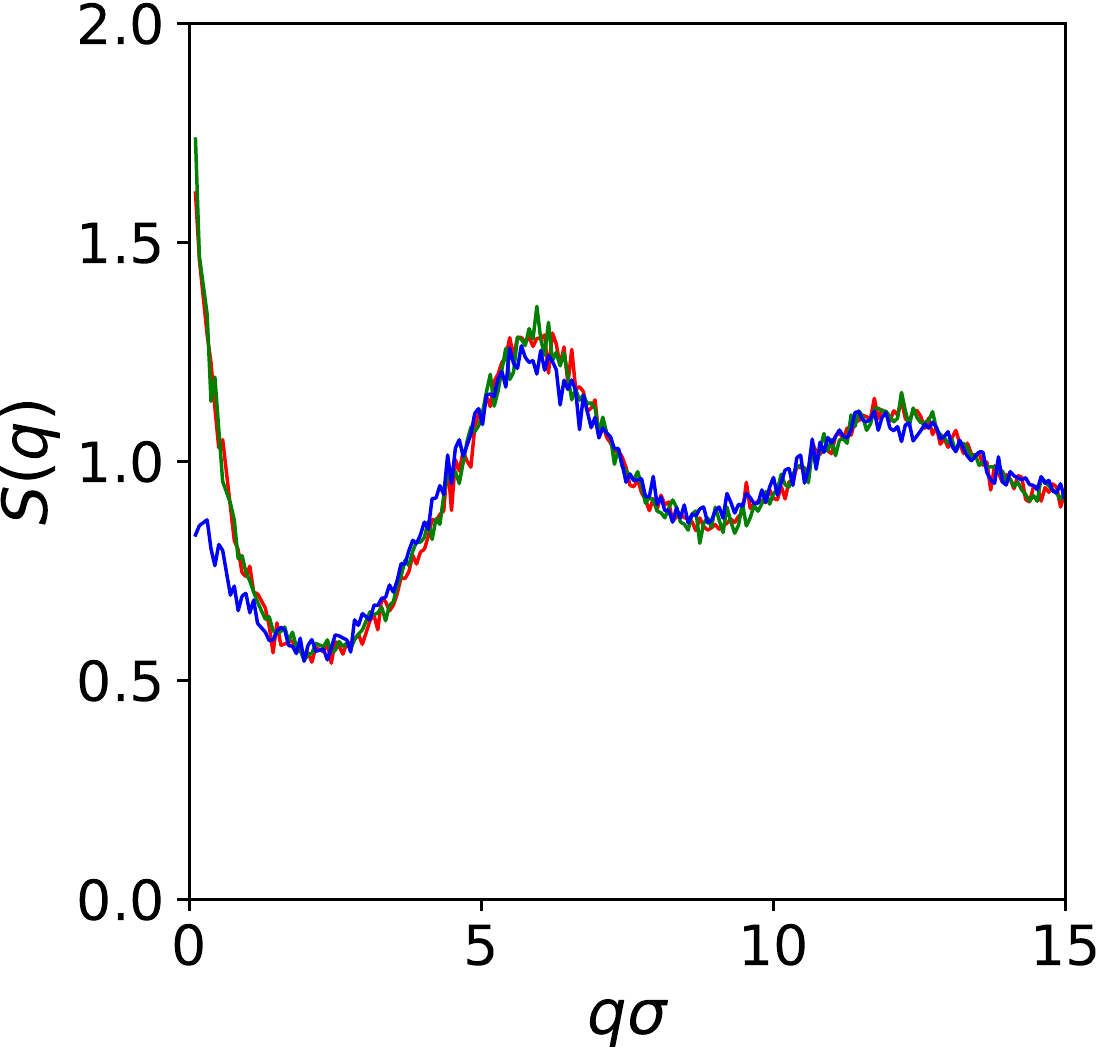}}}\hfill%
  \parbox[t]{.325\linewidth}{{\includegraphics[width=\linewidth]{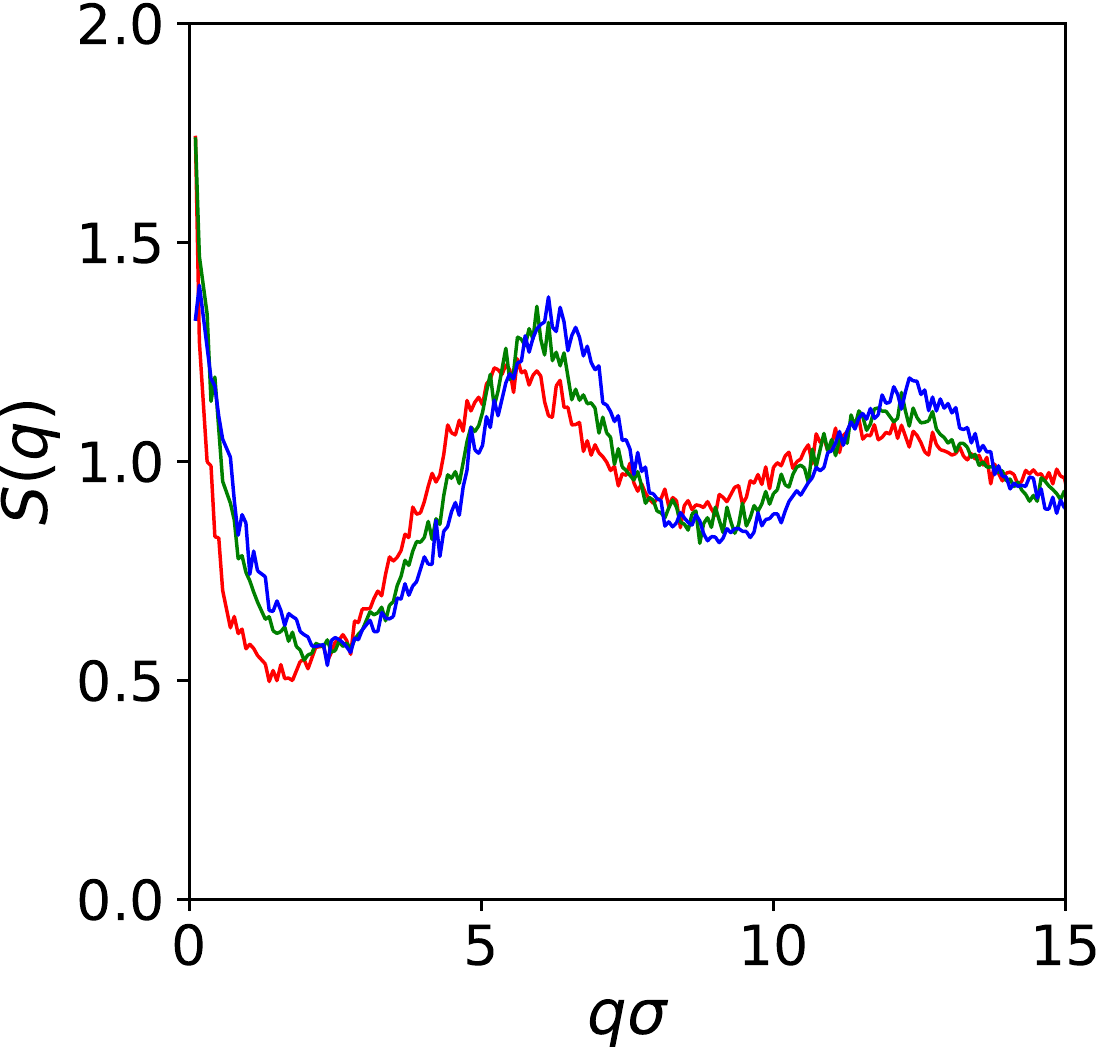}}}\hfill%
  \parbox[t]{.325\linewidth}{{\includegraphics[width=\linewidth]{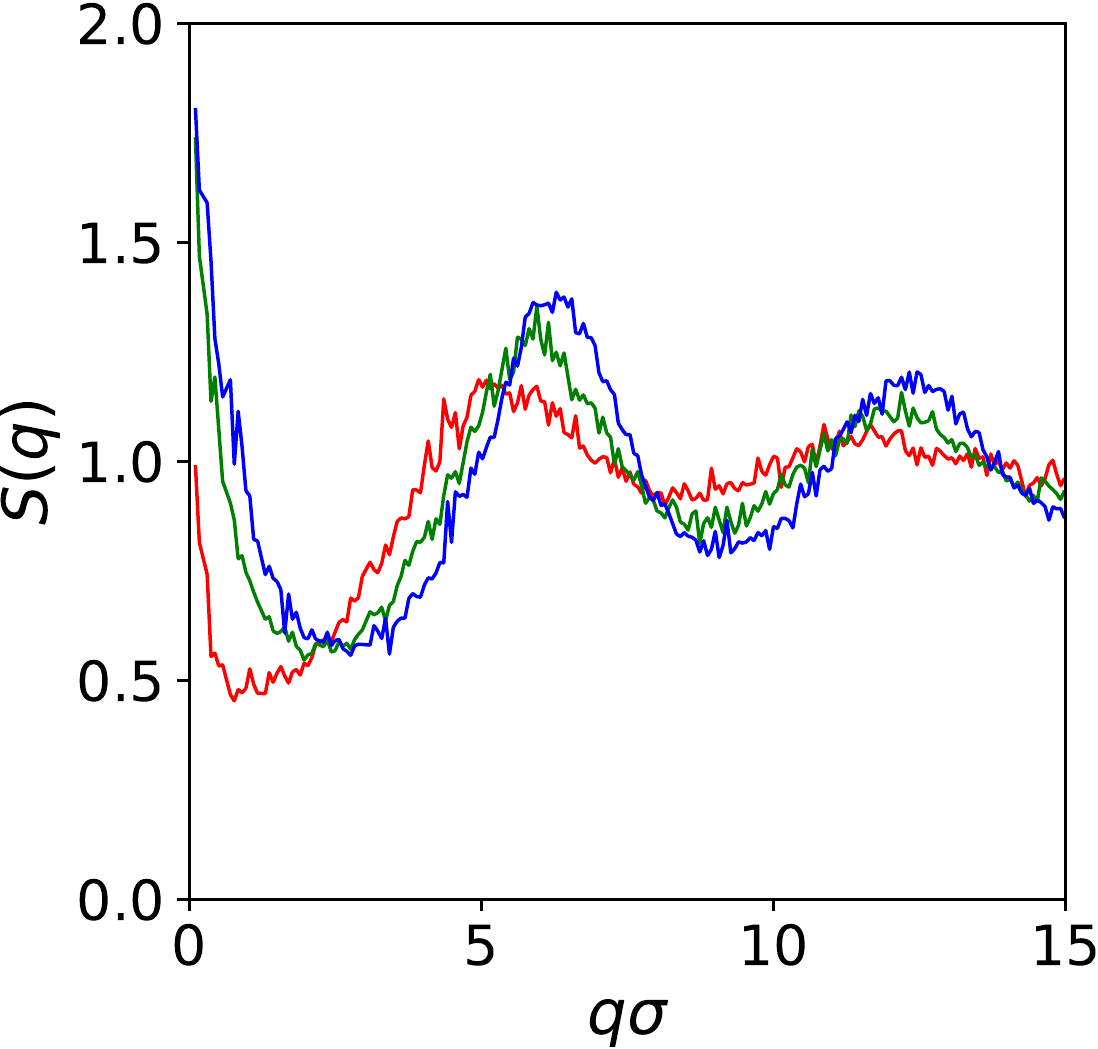}}}
}}
\caption{\label{fig:sq_lowdens_dr}
  Static structure factor of active hard-disks fluid at packing fraction
  $\eta=0.2$ with different rotational diffusion coefficients $D_r=0.1$, $1$,
  and $10$ (red, green, blue).
  Left: comparison at constant self-propulsion velocity
  (translational P\'eclet number) ${Pe}_t=v_0\sigma/D_t=10$;
  middle: comparison at constant P\'eclet number ${Pe}=v_0^2/D_rD_t=100$;
  right: comparison at constant persistence length
  (rotational P\'eclet number) ${Pe}_r^{-1}=v_0/\sigma D_r=10$.
}
\end{figure}

We thus compare the static structure factors obtained for the system
at moderate density, $\eta=0.2$, for various rotational diffusion
coefficients, while keeping one of the three parameters (${Pe}_t$,
${Pe}$, or ${Pe}_r^{-1}$) fixed. Varying $D_r$ over two orders of magnitude
shows (Fig.~\ref{fig:sq_lowdens_dr}) that the static structure factors
$S(q)$ are in fact independent on $D_r$ in the regime of moderately
strong self propulsion. The curves for different $D_r$ fall on top of each
other within error bars at fixed ${Pe}_t$, with the exception of a small
increase towards $q\to0$. The latter indicates that the appearance of
MIPS is influenced by both $v_0$ and $D_r$, while the dependence on $D_r$
has no significant structural precursors in the fluid.

At fixed ${Pe}$ (middle panel of Fig.~\ref{fig:sq_lowdens_dr}), an increase
in $D_r$ has a similar effect as increasing $v_0$ regarding the change
in nearest-neighbor structure that is reflected in the change of
the peak positions in $S(q)$. A similar conclusion holds for the evolution
of $S(q)$ with increasing $D_r$ at fixed persistence length (right panel
of the figure). Note that one expects the dynamics for $D_r\to\infty$
at fixed $v_0$
to become identical to that of passive hard spheres, because in this limit
the net effect of self propulsion vanishes. Figure~\ref{fig:sq_lowdens_dr}
suggests that this is true only if the limit is taken such that the
persistence length vanishes sufficiently quickly: only for the case
of fixed ${Pe}_t$, the structure factor approaches the passive one with
increasing $D_r$ at least at low $q$.

\begin{figure}
\indent{\parbox[t]{\linewidth}{
  \parbox[t]{.325\linewidth}{{\includegraphics[width=\linewidth]{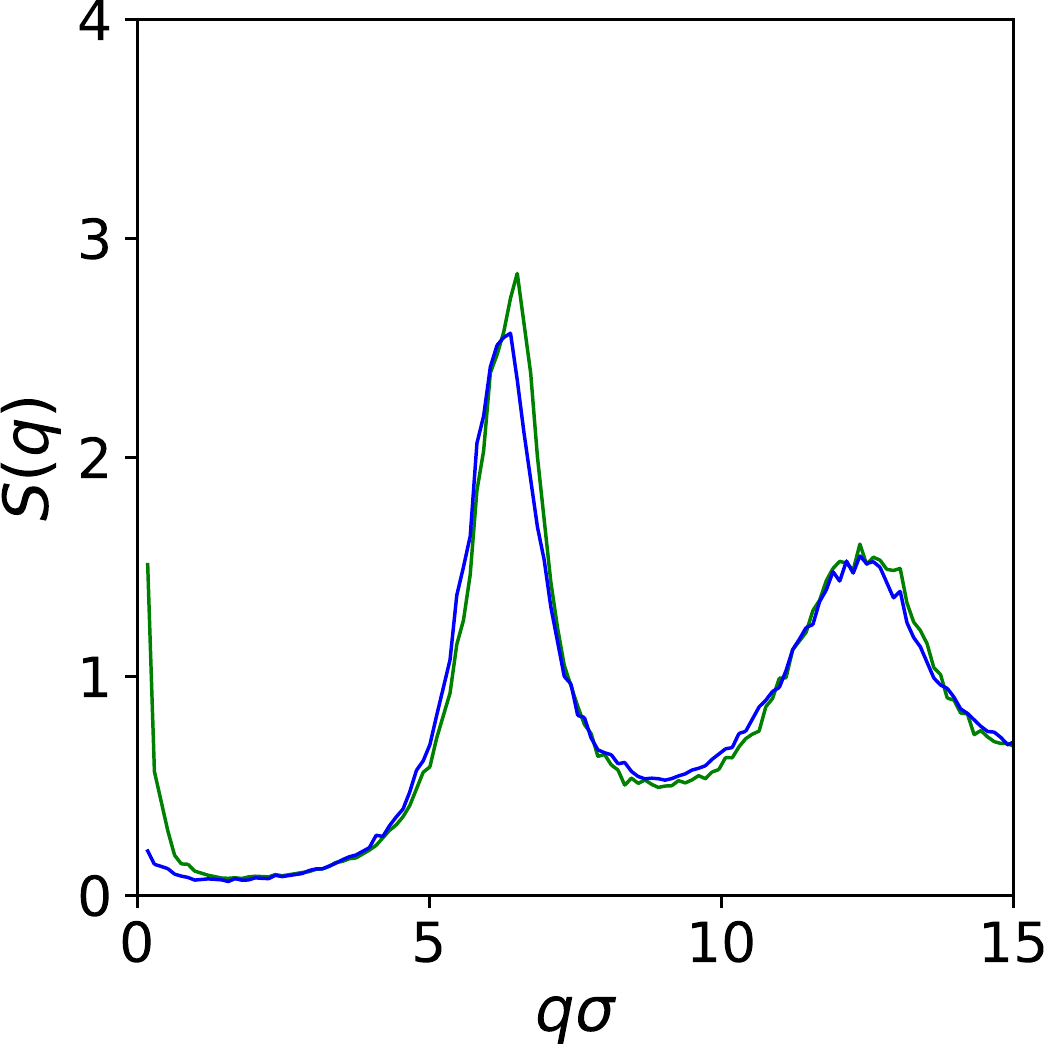}}}\hfill%
  \parbox[t]{.325\linewidth}{{\includegraphics[width=\linewidth]{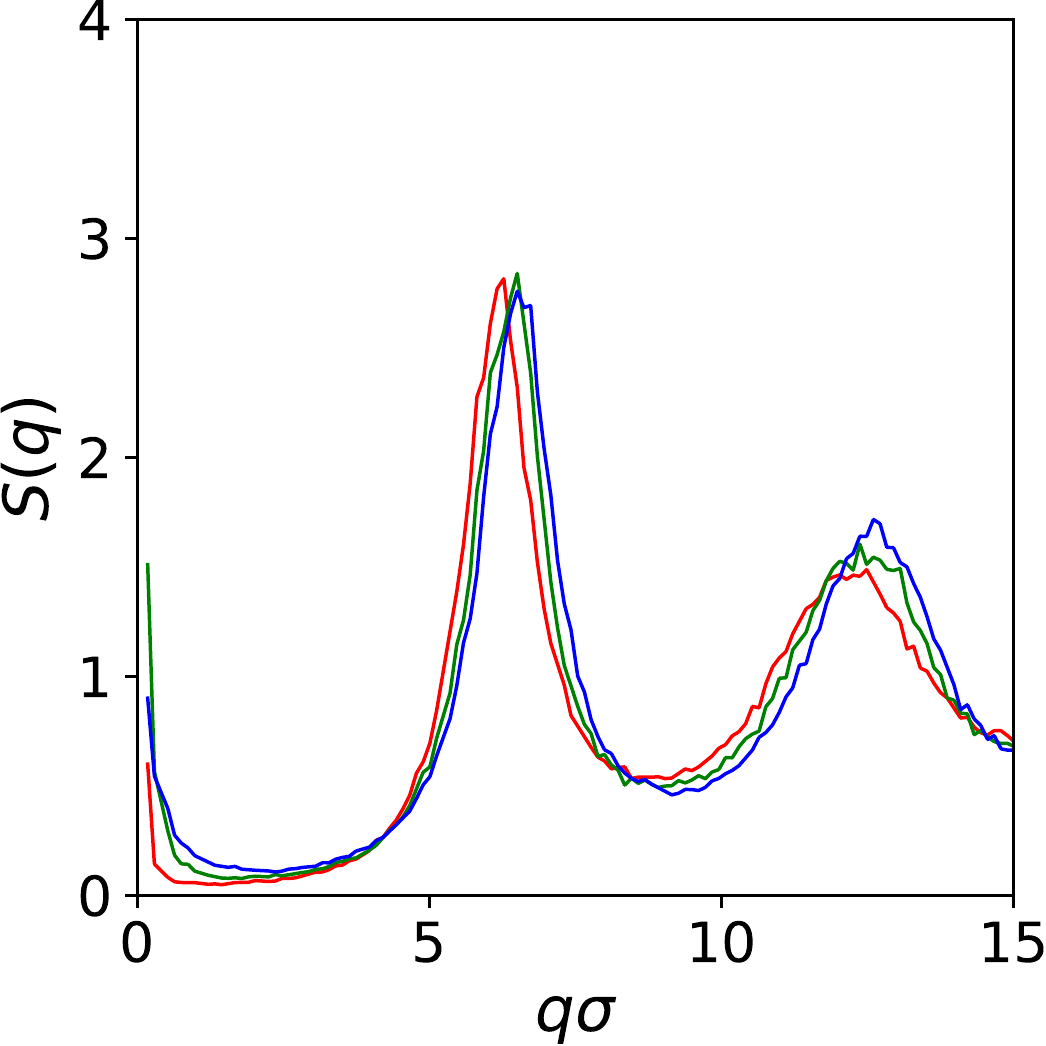}}}\hfill%
  \parbox[t]{.325\linewidth}{{\includegraphics[width=\linewidth]{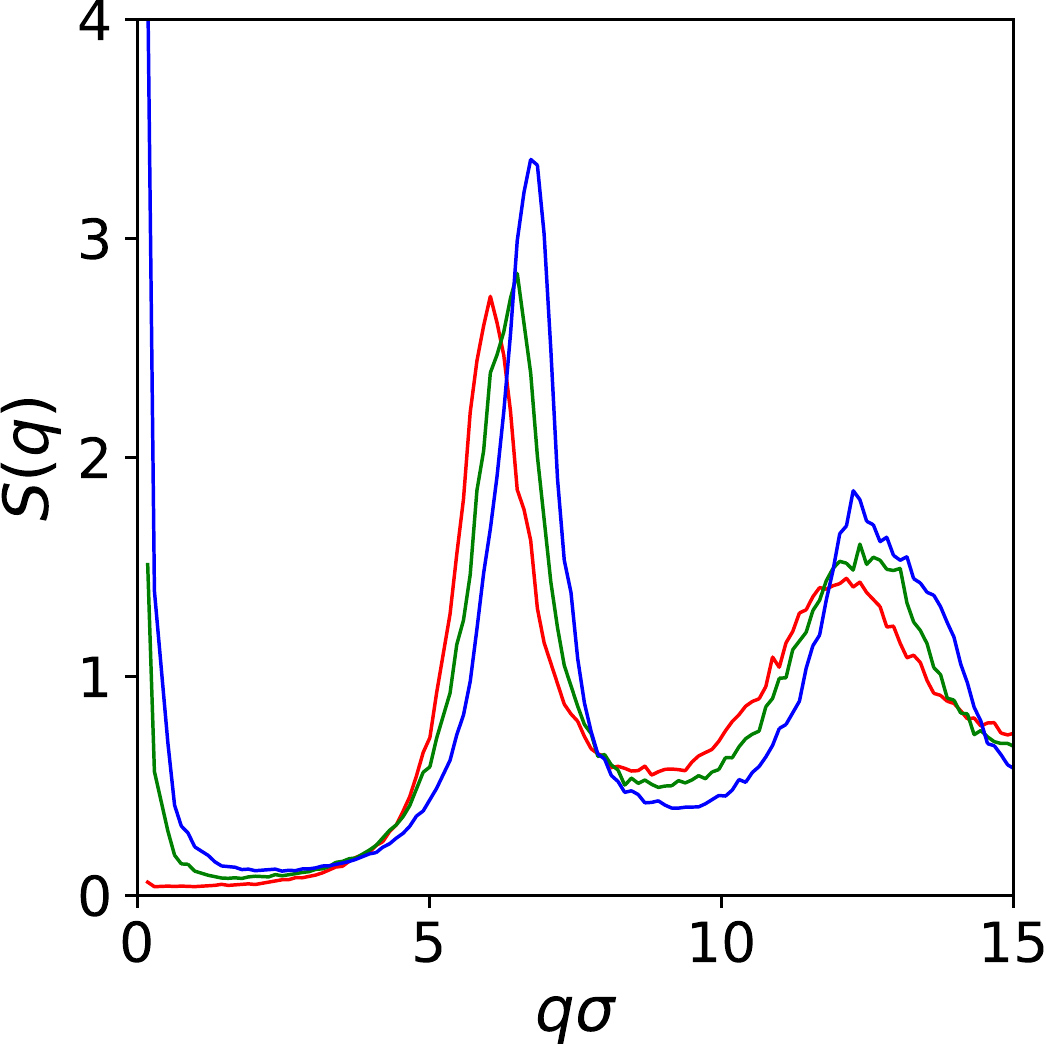}}}\\
  \parbox[t]{.325\linewidth}{{\includegraphics[width=\linewidth]{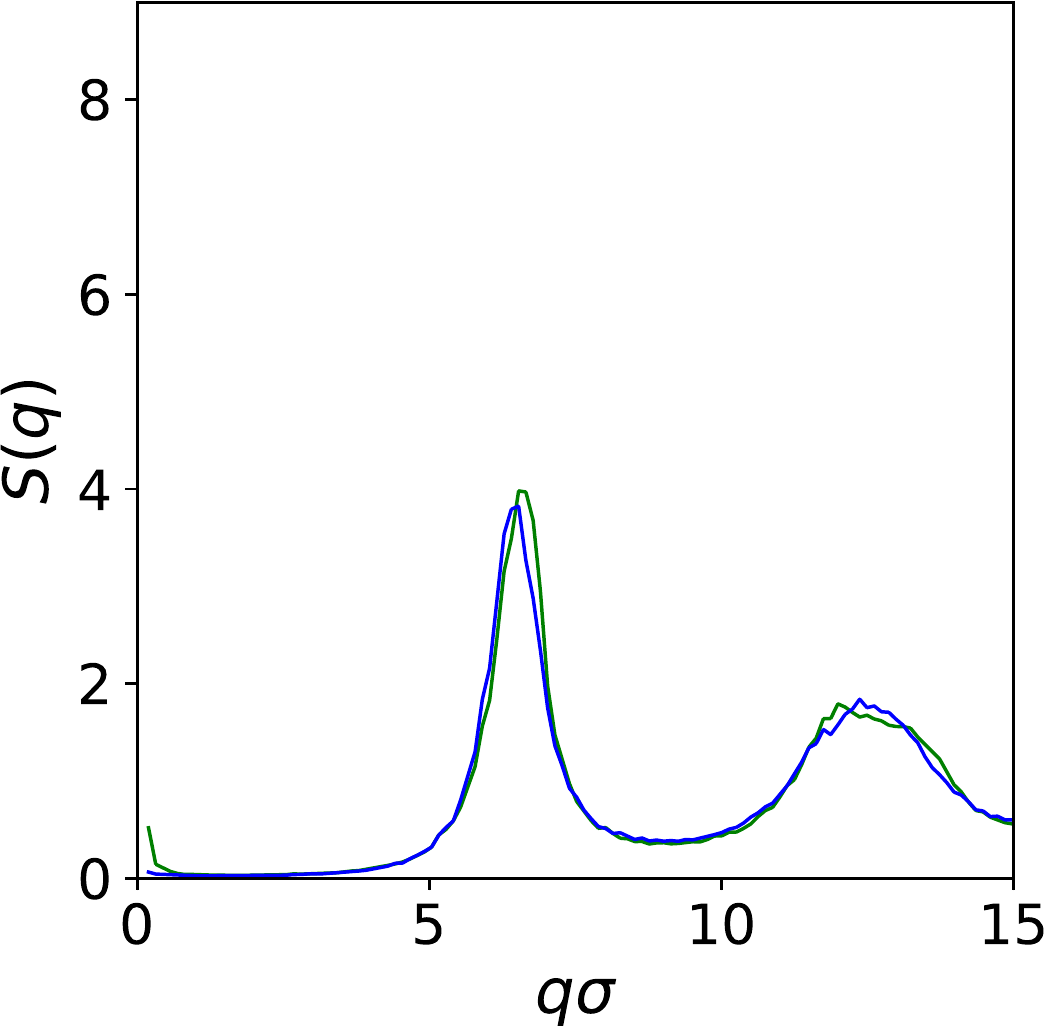}}}\hfill%
  \parbox[t]{.325\linewidth}{{\includegraphics[width=\linewidth]{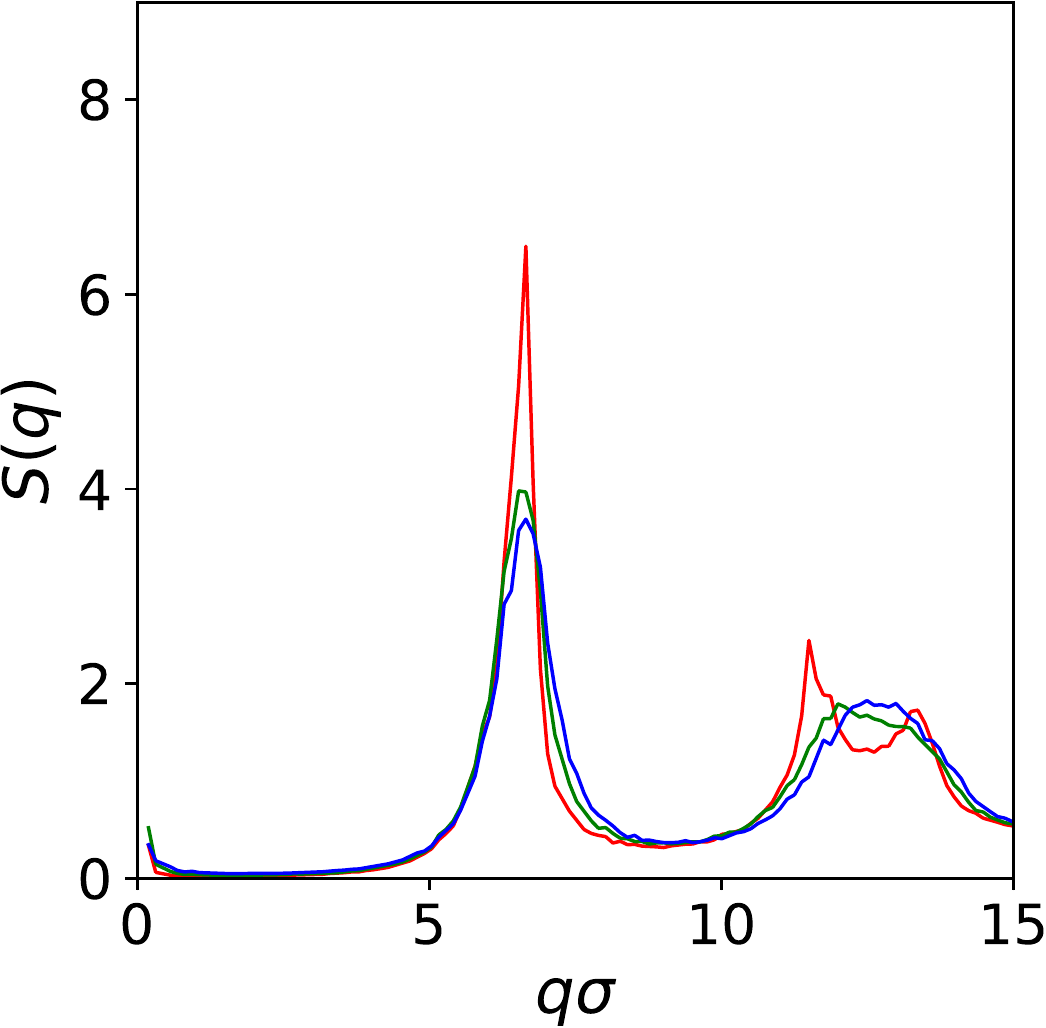}}}\hfill%
  \parbox[t]{.325\linewidth}{{\includegraphics[width=\linewidth]{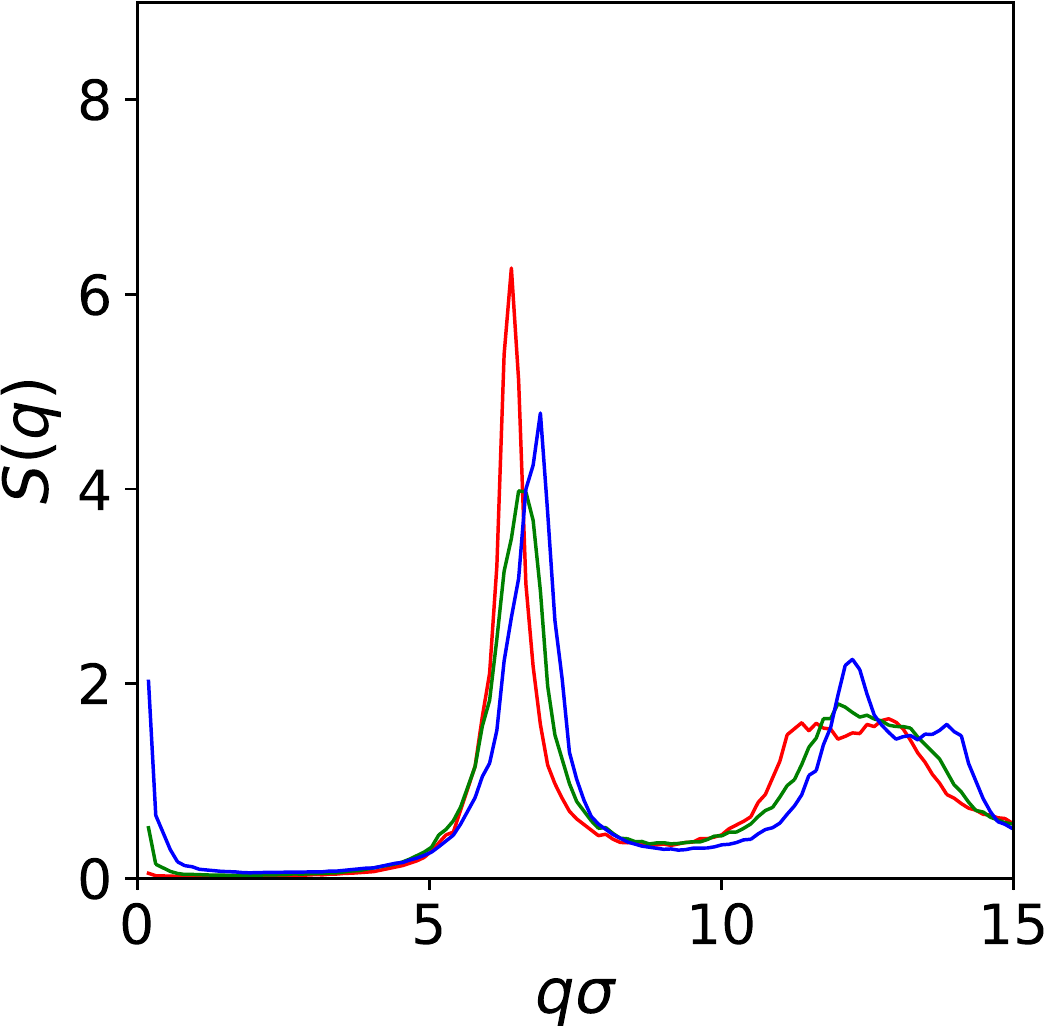}}}
}}
\caption{\label{fig:sq_highdens_dr}
  Static structure factor of active hard-disks fluid at packing fraction
  $\eta=0.6$ (top row) respectively $\eta=0.7$ (bottom row)
  with different rotational diffusion coefficients $D_r=0.1$, $1$,
  and $10$ (red, green, blue), at constant ${Pe}_t=10$ (left), constant ${Pe}=100$ (middle), and constant ${Pe}_r^{-1}=10$ (right).
  Results are omitted for the state point (${Pe}_t=10$, $D_r=0.1$), where
  the system is already phase separated.
}
\end{figure}

At high densities, the situation is less clear, because phase separation
sets in at very different self-propulsion velocities for different $D_r$.
Static structure factors for $\eta=0.6$ and $\eta=0.7$,
shown in Fig.~\ref{fig:sq_highdens_dr}, confirm the observation made above:
as long as the system remains in the homogeneous fluid state, $D_r$
itself appears to have very little influence on the static structure.
At fixed P\'eclet number or fixed persistence length, the increase of $D_r$
does not render the system more passive-like, but rather induces a change
in $S(q)$ that is similar to the one seen upon increasing $v_0$ at
fixed $D_r$.

\section{Conclusions}

We have obtained static structure factors and radial distribution functions
from simulations of active Brownian hard-disk fluids.
The system remains in the homogeneous fluid phase for all packing fractions
$\eta\lesssim0.7$ and low enough activities, $v_0\lesssim12\,D_t/\sigma$
for $D_r=1\,D_t/\sigma^2$; it also remains a homogeneous fluid for low
densities, $\eta\lesssim0.2$, and all the self-propulsion velocities we have
studied.

The evolution of the static structure factor with density, outside the
region of MIPS, resembles that of the passive fluid at finite $q$.
On the low-density side of the phase-separation spinodal, increasing
activity causes the fluid to exhibit slightly more pronounced ordering
concommittant with a shift of typical interparticle separations to smaller
distances. For the active hard-disk system, the static structure evolves
from the passive one to an essentially $v_0$-independent active one: the
$S(q)$ data with increasing $v_0$ converge to a well-defined limiting curve.

The high-density fluid reveals an interesting non-monotonic change in $S(q)$
with increasing $v_0$. Small self-propulsion velocities destroy the ordering
that is present in the passive dense fluid, but further increasing the
self-propulsion velocity reinstitutes more pronounced ordering at a shorter
average particle separation. This highlights the interplay of a
nearest-neighbor length scale (that determines the fluid structure in
equilibrium) and the length scale introduced by persistent swimming.

It is possible to qualitatively understand the
active-fluid $S(q)$ by analogy to the effects caused by an attractive
interaction in an equilibrium fluid. In particular the non-monotonic evolution
of $S(q)$ at high densities suggests an effective attraction of a range
that is around $10\%$ of the particle diameter. From this, an interplay
between the effective activity-induced attraction width and the interparticle
separation length arises. In the equilibrium fluid, such an interplay
causes non-monotonic changes in the dynamics that lead to a reentrant
melting of the glass \cite{Dawson2001,Pham2002}. Indeed, a similar reentrant
behavior of the glassy dynamics of an active fluid (albeit not an ABP fluid)
has been discussed \cite{szamel2015glassy}. It should, however, be stressed
that the effective-attraction description of $S(q)$ does not acknowledge
the non-equilibrium nature of the active fluid. While changes in
the glassy dynamics of an equilibrum fluid are largely governed by changes
in $S(q)$, the same need not be true for the ABP system.

The fact that activity enhances structural order in the low-density fluid might at first
sight seem surprising: The single-particle dynamics of ABP can be
mapped onto diffusion with an enhanced diffusivity, $D_\text{eff}\ge D_t$.
This mapping suggests a description in terms of an enhanced effective
temperature, and from such a mapping, one would expect the oscillations in
$S(q)$ to become less pronounced.
The qualitative mapping to an effective attraction, instead of an effective
temperature, much better explains the observed changes in $S(q)$.

The effect of rotational diffusivity on the low-density static structure
of the active hard-disk fluid is weak. At high densities, the influence
of $D_r$ is masked by the onset of phase separation, so that the trends
emerging in $S(q)$ are less clear. As already known from studies of MIPS,
it is not possible to maintain the high-density strongly active system
in a homogeneous fluid state. Doing so may be possible in suitably polydisperse
systems (either using size polydispersity or polydispersity in self-propulsion
speed).

Our data can be used as a reference for future theoretical studies which construct approximative closures for the structure based on the
Smoluchowski equation \cite{Haertel,SpeckEPL,FarageBraderArchive} 
or for mode-coupling-like dynamical theories that either need static
structure factor data as an input
\cite{FarageBraderArchive,Berthier2013,Szamel}
or can in principle approximate it based on the equilibrium one \cite{mct}.
As a remark, however, the extension of theoretical frameworks to calculate
$S(q)$ and related structural quantities that characterize the
non-equilibrium steady state of active particles is not obvious, even for radially-symmetric pair potentials. 
There are two basic reasons for that: first, active Brownian particles possess an internal orientational degree of freedom
which represents the direction of their intrinsic motion. This orientational degree of freedom is irrelevant for spherical
passive systems but similar to molecular liquids with rotational symmetric shape (such as rods).
For these liquid-crystalline systems, it is much more difficult to derive and numerically solve
integral equations closures to access the orientationally averaged structure factor for the particle centers
\cite{Gubbins,Patey,Klapp}. Second, more importantly, active Brownian system are not in equilibrium which brings about
complications in the description of the steady state.

\ack{
We gratefully acknowledge funding from Deutsche Forschungsgemeinschaft (DFG)
within the Special Priority Programme SPP~1726 ``Microswimmers'',
grants VO~1270/7-2 and LO~418/17-2.
}

\section*{References}

\bibliography{lit}
\bibliographystyle{iopart-num}

\end{document}